\documentstyle[12pt,amstex,epsf,epsfig,subfigure,float]{article}  
\begin{document}

\makeatletter
\long\def\@makecaption#1#2{
 \vskip 10pt
 \setbox\@tempboxa\hbox{{\bf #1: }{\small \em #2}}
 \ifdim \wd\@tempboxa >\hsize \unhbox\@tempboxa\par \else \hbox
to\hsize{\hfil\box\@tempboxa\hfil}
 \fi}
\makeatother

\renewcommand\topfraction{1.}
\renewcommand\bottomfraction{1.}
\renewcommand\textfraction{.0}


\begin{titlepage}

\pagenumbering{arabic}
\vspace*{-1.5cm}
\begin{tabular*}{15.cm}{lc@{\extracolsep{\fill}}r}
{\bf DELPHI Collaboration} & 
\hspace*{1.3cm}
\epsfig{figure=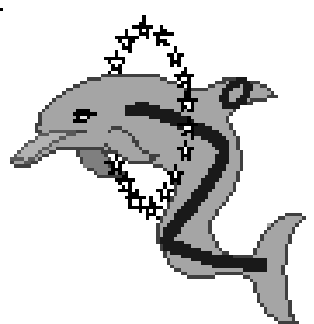,width=1.2cm,height=1.2cm}
&
DELPHI 99-150 RICH 95
\\
& &
03 October, 1999
\\
&&\\ \hline
\end{tabular*}
\vspace*{2.cm}
\begin{center}
\Large 
{\bf \boldmath
MACRIB \\ 
High efficiency - high purity \\ 
hadron identification for DELPHI
} \\
\vspace*{2.cm}
\normalsize { 
   {\bf Z. Albrecht, M. Feindt and M. Moch}\\
   {\footnotesize Institut f{\"u}r Experimentelle Kernphysik \\
    Universit{\"a}t Karlsruhe}\\
   
}
\end{center}
\vspace{\fill}
\begin{abstract}
\noindent
Analysis of the data shows that hadron tags of the two 
standard DELPHI particle identification packages RIBMEAN 
and HADSIGN are weakly correlated. This led to the idea 
of constructing a neural network for both kaon and proton 
identification using as input the existing tags from 
RIBMEAN and HADSIGN, as well as preproccessed TPC and RICH 
detector measurements together with additional dE/dx 
information from the DELPHI vertex detector. It will be 
shown in this note that the net output is much more 
efficient at the same purity than the HADSIGN or 
RIBMEAN tags alone. We present an easy-to-use routine 
performing the necessary calculations.

\end{abstract}
\vspace{\fill}

\vspace{\fill}
\end{titlepage}







\section{Introduction}
\begin{figure}[H]
  \centering
  \epsfig{file=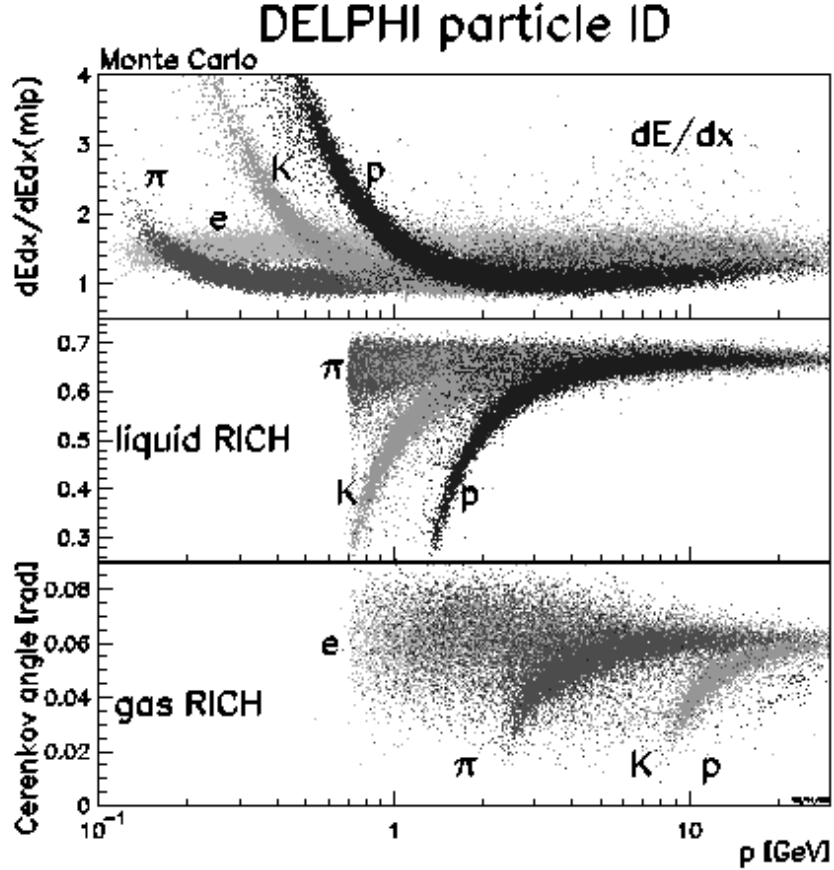,width=11cm,clip=,bbllx=72,bblly=200,bburx=484,bbury=637}
  \caption {DELPHI particle identification}
  \label{fig:pid}
\end{figure}

%
With the Rich Imaging Cherenkov Chamber (RICH) DELPHI possesses a unique and
very important hadron identification tool. The two informations from the
liquid and gas radiators can be combined with the dE/dx information
measured in the Time Projection Chamber (TPC) \cite{combined} to achieve
separation of hadron species at different momenta (fig.~\ref{fig:pid}).
In different momentum ranges the single detectors (TPC, RICH) can
distinguish the hadrons in different ways and quality. The description
of the particle identification algorithms for the Rich detectors based
on RIBMEAN\cite{rib}. On first sight, there is no problem
if the bands of the single hadrons can be well separated. An obvious problem
is how to deal with momentum ranges where one detector cannot separate
between different particle hypotheses and another detector is in the
so-called veto mode i.e. when one expects no signal for a track in the
detector. This happens e.g. in the range $5~\textrm{GeV}<p<10~\textrm{GeV}$
for kaons and protons. This is handled in HADSIGN and RIBMEAN in different
ways \cite{RICH,HAD}. Veto signals are especially hard to establish in the
case of background due to large particle density (in case of RIBMEAN see
\cite{veto}). Especially the background photon handling is done differently
in HADSIGN and RIBMEAN. We later show that even in the case where two bands
are clearly separated a problem appears in at least RIBMEAN. We do
not have intermediate information to check HADSIGN at the SDST level. From
the above considerations it is clear that the construction of simple tags is
strongly momentum dependent and not a trivial task.

\begin{figure}[t]
\begin{center}
\leavevmode 
\includegraphics[bb=0 99 590 643, width=0.6\textwidth]
                {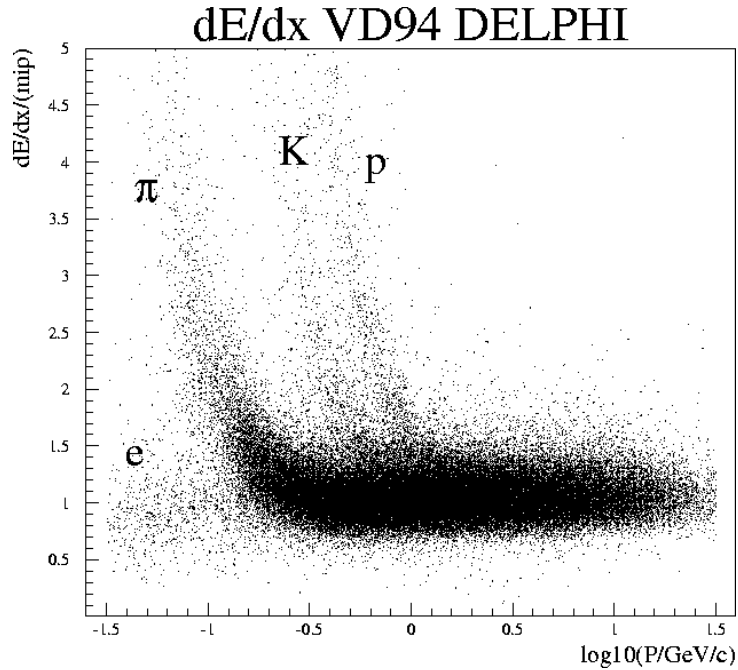}
\caption[dE/dx measurement with the vertex detector.]
        {\label{vddedx} The vertex detector dE/dx with three hits 
                        and more for the whole
                        momentum range, measured with the 94 
                        vertex detector. Clearly visible are 
                        the e, $\pi$, Kaon and proton bands.}
\end{center}
\end{figure}

As one can see in figure \ref{fig:pid}, below 0.7 GeV no RICH
information is available. In addition, 21\% of the tracks have less then 30 TPC wire hits,
so that in these cases no particle identification is possible with the TPC
as the only detector.
As in the case of higher momentum particles, i.e. above 0.7 GeV, an independent
measurement would contribute a large boost to the efficiency of hadron tagging. This
is supplied by the Vertex Detector, usually used for spatial measurements only,
but the signal in the silicon tracker also gives signal height information
proportional to the deposited energy lost by the passing particle. The signals
from the different hypothesis can be seen in figure \ref{vddedx}.

In this note we present the MACRIB\footnote{MACRIB=``Michael And 
Company's Rich Identification Backage''. The local Karlsruhe dialect cannot
distinguish between P and B.} package
which combines the benefits of both RICH algorithms and dE/dx information
for the high momentum region. For the low momentum region the combined dE/dx
information of both the TPC and the VD are used. The combination is done using
a neural network, which has a much better performance than any of it's input
variables alone. 
 
%
\section{Comparison of HADSIGN and RIBMEAN}
\begin{figure}[t]
  \centering
  \mbox{
  \subfigure{\epsfig{file=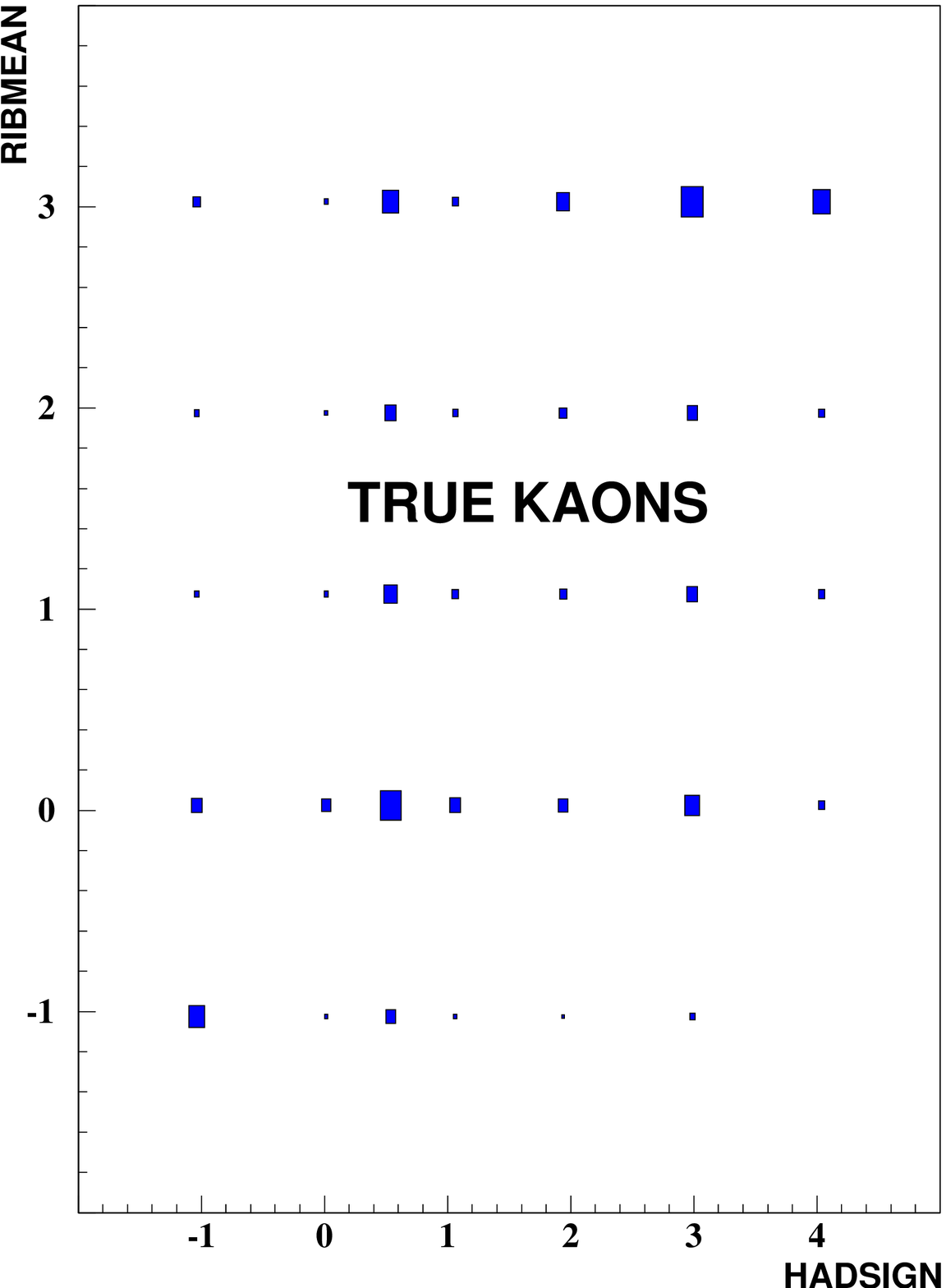,
            width=7cm}}
  \subfigure{\epsfig{file=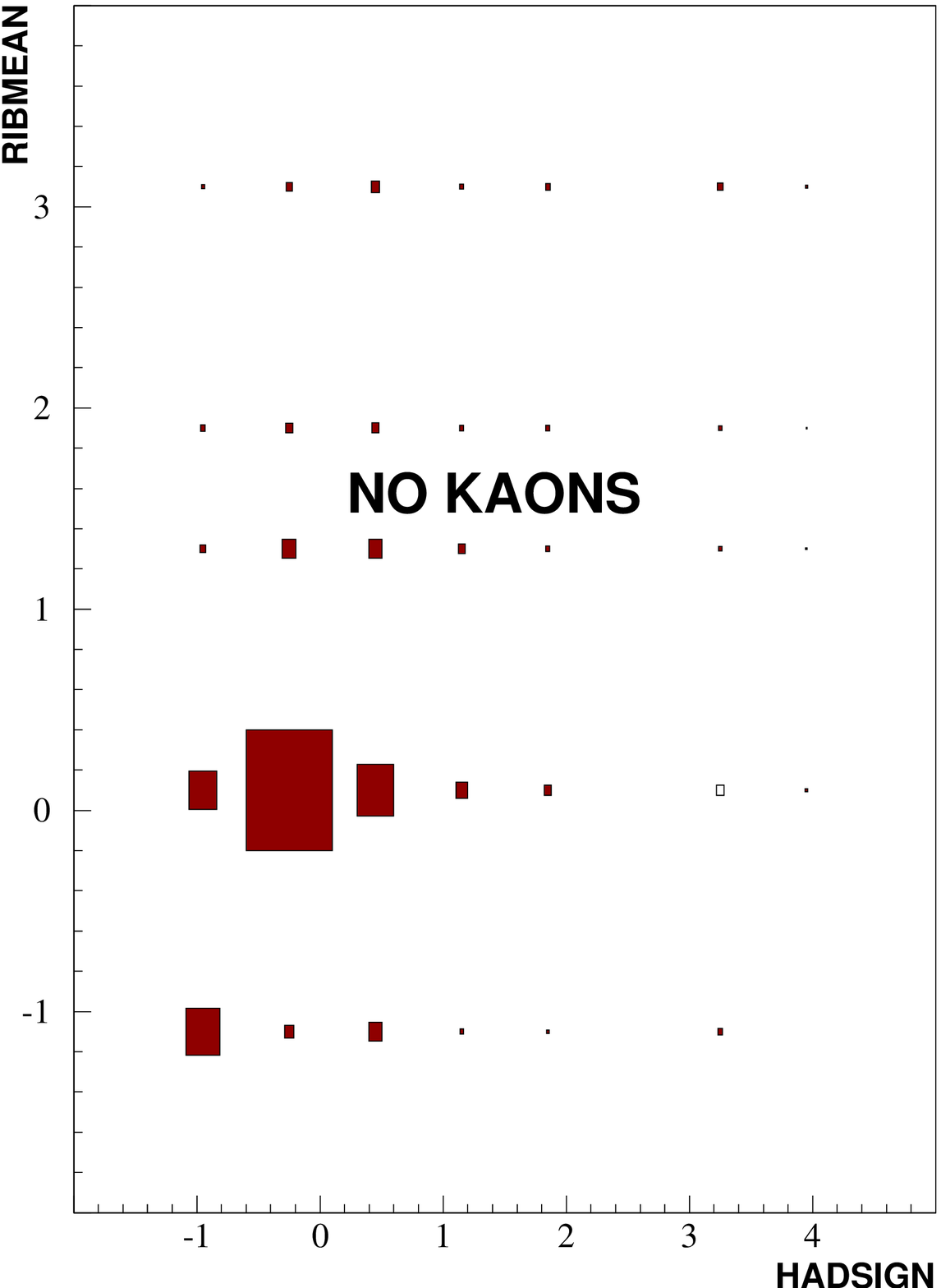,
            width=7cm}}
  }
  \caption[Comparison of kaon tags]
          {Comparison of combined the HADSIGN and RIBMEAN kaon tags shows the
           anticorrelation in the tagging of the same particle as a tight
           kaon through one routine and as a non-kaon through the other.}
  \label{fig:kaontags}
\end{figure}

%
RIBMEAN provides a tag between -1 and 3 according to the probability for a given
particle calculated within the routine. -1 means that no information is available, 0
stands for sure background and 1,2,3 for the loose, standard and tight kaon tags. In
addition to RIBMEAN, HADSIGN introduces a very tight kaon tag with code 4 and a
very loose tag with code 0.5. For the case of no information and no identification both
algorithms give -1 and 0 respectively.
In the ideal case both approaches should give the same or
at least similar tags for a particle. A comparison of the tags shows that they are
hardly correlated (see fig~\ref{fig:kaontags}). There are situations where RIBMEAN
works more efficiently and vice versa. A possible reason for this discrepancy is
the different approach for the ring finding algorithm in the RICH detectors.
RIBMEAN divides the plane around a track intersection point into rings with different
radii according to each particle hypothesis ($e,\ \mu,\ \pi, K$ and $p$). The
photoelectrons are weighted within the rings. Different weights are given depending on
the background environment, e.g. on the (non-)ambiguity and the geometrical position in
the ring. The cluster with the largest sum of weights is selected. For isolated tracks
this algorithm gives a very reliable result. For high density tracks the rings overlap
and not all ambiguities can be solved. In these cases the performance of this method drops.
In contrast, HADSIGN treats background by including it in the model. It sums up the
photo-electrons as a function of the distance from the intersection point, regardless
of its origin. Next a maximum likelihood technique is applied to the obtained distribution.
According to the result of the fit HADSIGN assigns a probability to the different
hypotheses. In this way, the performance of HADSIGN is relatively independent of background.
%
%
\section{A neural network approach}
\begin{figure}[t]
  \centering
  \epsfig{file=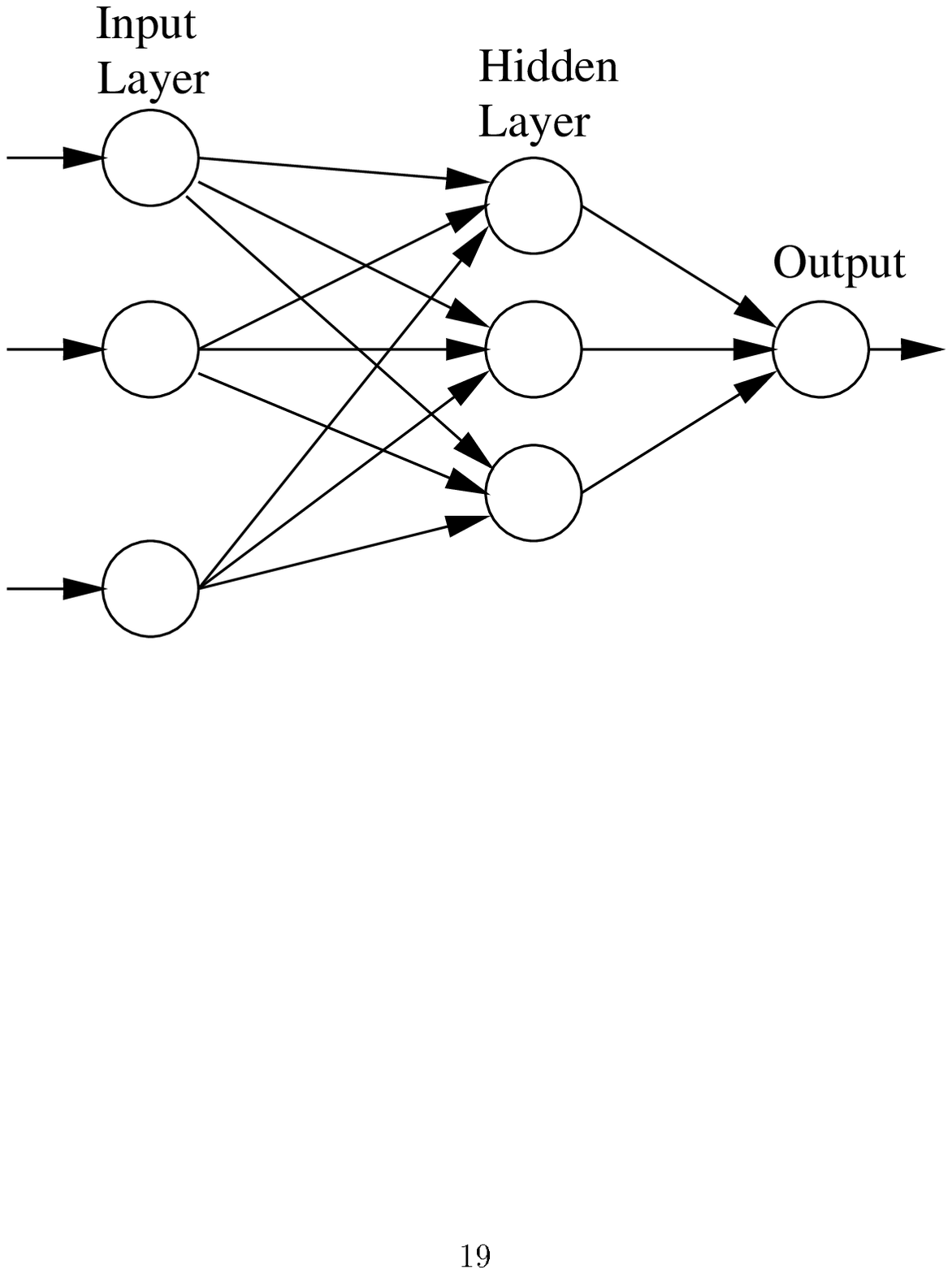,width=7cm,
          clip=,bbllx=100,bblly=250,bburx=460,bbury=480}
  \caption {An example for a simple feed forward neural network with three layers.}
  \label{fig:nn}
\end{figure}

%
This  observation was the motivation for optimally combining both algorithms using
a simple feed forward network with backpropagation. A schematic example of the
topology of this kind of network is given in figure \ref{fig:nn}. Three sets of neural
networks have been constructed for kaon and proton identification:
\begin{itemize}
  \item $p>0.7$~GeV with full RICH information,
  \item $p>0.7$~GeV with no liquid RICH information,
  \item $p\leq 0.7$~GeV TPC dE/dx combined with dE/dx measurements of the vertex
        detector.
\end{itemize}
The distinction between the cases with and without liquid information is necessary for
the Monte Carlo Data agreement was not satisfactory. The reason originates from the fact that there are periods
in the data where the RICH is only partly functional. For the case of no liquid RICH
information present, a new network is required due to the relatively large number of input nodes
requiring liquid information and to the lack of non-liquid radiator information in the 1994
Monte Carlo simulation for the training. For the training of events with only gas information
a MC sample from 1993 has been used.
\subsection{Input variables for the nets with liquid RICH information}
The networks have four layers, 16 input nodes, one bias node, 19 nodes in the
hidden layer and one output node. Several variables of the RICH and the TPC are
used as input variables, as detailed below. Many of the variables have been
optimised for the use as a neural network input. For the training we used 800,000
kaons and approximately the same number of non-kaons from the 94C2
$ Z^{0} \to q\bar{q}$ Monte Carlo
datasets. We use the following variables as input for the kaon net:
\begin{enumerate}
\item the {\bf RIBMEAN RICH kaon tag}. The order of codes 0 and -1 has been
      interchanged in order to make the kaon probability a monotonically rising 
      function of the tag.
\item the {\bf HADSIGN RICH kaon tag}. The order of codes 0 and -1 has been
      interchanged in order to make the kaon probability a monotonically rising 
      function of the tag.
\item the {\bf dE/dx kaon tag} from HADSIGN. The order of codes 
      0 and -1 has been
      interchanged in order to make the kaon probability a monotonically rising 
      function of the tag.  
\item the logarithm of the ratio of the {\bf gas RICH probabilities} of 
      the {\bf kaon} hypothesis to the {\bf pion} hypothesis for kaon 
      separation from pions with gas RICH. If no gas measurement is available,
      the value is set to 0.
\item the logarithm of the ratio of the {\bf liquid RICH probabilities} of 
      the {\bf kaon} hypothesis to the {\bf pion} hypothesis for kaon 
      separation from pions with liquid RICH. If no liquid measurement is
      available, the value is set to 0.
\item the {\bf electron neural net output}  from ELEPHANT, to
      suppress electrons faking kaons.
\item the {\bf combined HADSIGN proton tag}, to suppress protons
      faking kaons. The order of codes 0 and -1 has been
      interchanged in order to make the proton probability a monotonically rising 
      function of the tag.
\item the logarithm of the ratio of the {\bf dE/dx probabilities} of 
      the {\bf kaon} hypothesis to the {\bf pion} hypothesis for kaon 
      separation from
      pions with dE/dx. If no dE/dx measurement is available, the value is
      set to 0.
\item the logarithm of the {\bf likelihood ratio} for the 
      {\bf number of photons} in the
      {\bf liquid} RICH for the {\bf kaon} and {\bf pion} hypotheses. 
      This is computed 
      in the following way:
      The number of reconstructed photons in the liquid radiator is 
      subtracted from the expected number of photons for the pion hypothesis as
      a function of the expectation and the particle hypothesis, 
      in 7 expectation bins from 5 to 20 photons. 
      In each of the bins the normalised Monte Carlo distribution is fitted
      to a double Gaussian.    
      The variation of the 6 fit parameters in the 7 expectation bins is
      then parametrised, such that the probability density is available as 
      a continuous function of expectation and particle hypothesis.
      The input variable is the logarithm of the ratio of these functions
      for kaon and pion hypotheses.
\begin{figure}[t]
  \centering
  \includegraphics[bb=40 184 520 645, width=7cm]{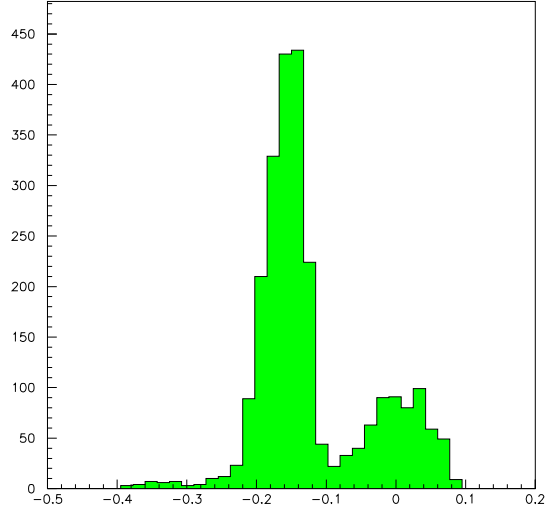}
  \caption[Liquid RICH angle distribution]
          {Performance of the liquid RICH data on the \v{C}erenkov angle. The plot
           shows the difference between measured and expected angle for the
           pion hypothesis for true kaons in the momentum region 0.7-0.9 GeV, i.e. in a
           region where the kaon band is significantly separated from the pion band. The
           distribution around -0.16 shows correctly identified kaons, whereas the peak at 0.
           shows misidentified kaons.}
  \label{fig:liqangle}
\end{figure}
\item the logarithm of the {\bf likelihood ratio} for the 
      {\bf number of photons} in the
      {\bf liquid} RICH for the {\bf kaon} and {\bf proton} hypotheses 
      to suppress proton contamination.          
\item the logarithm of the {\bf likelihood ratio} for the 
      {\bf number of photons} in the
      {\bf gas} RICH for the {\bf kaon} and {\bf pion} hypotheses 
      in the kaon gas band region.

\item the logarithm of the {\bf likelihood ratio} for {\bf kaon} 
      and {\bf pion} hypotheses
      in the {\bf kaon gas veto} region. This is computed in the following way:
      In the momentum region where one expects to see a ring for pions, but
      not for kaons, we distinguish the following cases:
      \begin{itemize}
      \item no photon is observed (i.e. a good veto): 
          as a function of the expected number of photons for the pion 
          hypothesis, and separately for tracks including the Outer Detector
          (signalling a more reliable track interpolation in the RICH)
          the ratio of real kaons to all tracks as expected 
          is parametrised.
      \item one photon is observed (i.e. a slightly worse veto):
          this one photon could be real or background, thus the separation 
          is worse in this case. The same procedure as above is performed.   
      \item more than one photon is observed, and an angle measurement performed
          (i.e. a bad or no veto): Even in this case one can distinguish 
          how often such an assignment is wrong. Both the angle and the 
          number of photon measurements should be compatible with the pion
          hypothesis.    
      \end{itemize}  
      Outside the interesting momentum region the variable
      is set to 0.5.

\item the information if the {\bf OD} was used in the track reconstruction.
\item the {\bf muon tag} for the kaon separation from muons. 
\item the momentum of the particle
\begin{figure}[t]
  \centering
  \epsfig{file=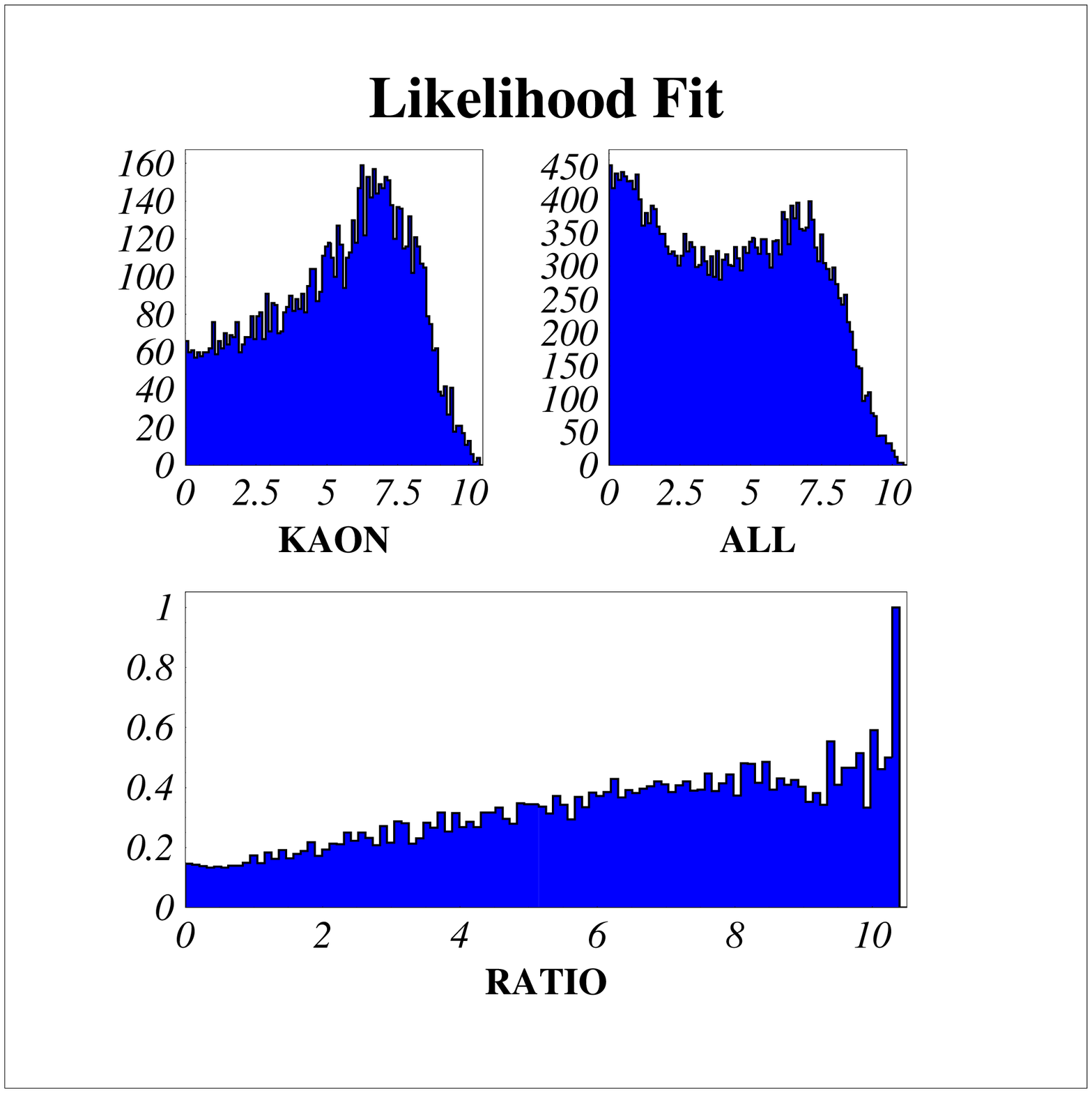,
          width=8cm,clip=,bbllx=76,bblly=190,bburx=485,bbury=665}
  \caption[Likelihood ratio for no or only one photon]
          {In the momentum region where one expects to see a ring for pions, but
           not for kaons for the case when no or only one photon is observed the following
           likelihood fit is made. As a function of the expected number of photons for the pion 
           hypothesis the ratio of real kaons to all tracks as expected 
           is parametrised.}
  \label{fig:liklyfit}
\end{figure}

\item the logarithm of the {\bf likelihood ratio} for the 
      {\bf \v{C}erenkov angle measurement} in the
      {\bf gas} RICH for the {\bf kaon} and {\bf pion} hypotheses 
      in the kaon gas band region.

\item the logarithm of the {\bf likelihood ratio} for the 
      {\bf \v{C}erenkov angle measurement} in the
      {\bf liquid} RICH for the {\bf kaon} and {\bf pion} hypotheses 
      in the kaon liquid band region below 4 GeV. Outside this region
      the variable is set to 0.5, outside the usual region on the background 
      side.
\item the probability of the {\bf dE/dx pull} of the {\bf kaon} hypothesis.      
\end{enumerate}
In order to reduce the number of input variables, two pairs of RICH measurements
were put together to form one without losing information, since the momentum
regions of the pairs do not interfere. These are the likelihood ratios for the
number of photons in the liquid and gas radiators and the likelihood ratios for
the Cerenkov angle measurement for the liquid and gas.

For the proton network an analogous set of input variables were used and for the
training $\approx$250,000 protons were used mixed with the same amount
of background. Additionally
a pion veto region is introduced for the liquid radiator in the momentum region
between $0.7~\text{GeV}<p<1.2~\text{GeV}$. The information is treated in a similar way as is
done for the kaons in the gas veto region.
\subsection{Net input variables without liquid RICH information}
The kaon network has three layers, 16 input nodes, one bias node, 17 nodes in the
hidden layer and one output node. RICH gas
and TPC variables are used as input variables. For the training we used the 93D2
$q\bar{q}$ simulation with a sample of 130,000 kaons and approximately the same number of
non-kaons. We use the following variables as input for the kaon net (for
the detailed description of the variables see section before):
\begin{enumerate}
\item the {\bf RIBMEAN RICH kaon tag}.
\item the {\bf HADSIGN RICH kaon tag}.
\item the {\bf HADSIGN combined proton tag}
\item the {\bf dE/dx kaon tag} from HADSIGN.  
\item $\log(\frac{P_K}{P_\pi})$ from the {\bf TPC dE/dx} measurement
\item {\bf Prob(Kaon Pull)} from the TPC.
\item {\bf Prob(Pion Pull)} from the TPC.
\item $\log(\frac{P_K}{P_\pi})$ from the {\bf gas RICH}
\item $\log\text{-likelihood}\left(\frac{N\gamma_{K}}{N\gamma_{\pi}}\right)$ in the
  {\bf kaon gas band} region.
\item $\log\text{-likelihood}\left(\frac{\Theta_{K}}{\Theta_{\pi}}\right)$
  in the {\bf kaon gas band} region.
\item for $0~\text{or}~1\gamma: \frac{K}{all}(N\gamma_{\pi,expected})$ or\\
  for $\geq2\gamma: \text{Prob}(\Theta_\pi,N\gamma_\pi)$ with 2 degrees of freedom in
  the {\bf kaon gas veto} region.
\item the logarithm of the ratio of the {\bf dE/dx probabilities} of 
      the {\bf kaon} hypothesis to the {\bf pion} hypothesis for kaon 
      separation from pions with the {\bf vertex detector}.
\item the logarithm of the ratio of the {\bf dE/dx probabilities} of 
      the {\bf proton} hypothesis to the {\bf pion} hypothesis for heavy particle 
      separation from pions with the {\bf vertex detector}.
\item the {\bf electron neural network output}
\item the {\bf muon tag}
\item the {\bf OD} flag
\item the {\bf momentum} of the particle
\end{enumerate}
To partially compensate the loss of the liquid radiator information in the momentum region
of $0.7~\text{Gev}<p<2.5~\text{Gev}$, the vertex detector measurement was introduced into the
net. The separation performance of the VD dE/dx in this region is not as good as of the
liquid RICH but it helps to solve some ambiguities. For the proton network $\approx$50,000 protons
were used mixed with the same amount of background and all kaon inputs were changed to proton
variables. The proton net has 15 input nodes and one bias node, 17 nodes in the hidden layer
and 1 output node.
\subsection{The low momentum neural net}
The kaon network has three layers, 9 input nodes, one bias node, 9 nodes in the hidden layer
and one output node. For the input variables a composition of TPC and VD information has been
used. For the training we used the 94C2 $q\bar{q}$ simulation with a sample of $\sim$5,000 kaons
and approximately the same number of non-kaons. We use the following variables as input for
the kaon net:
\begin{enumerate}
\item $\log(\frac{P_K}{P_\pi})$ from the {\bf TPC dE/dx} measurement. If no TPC information
  is available the input is set to 0.
\item $\log(\frac{P_K}{P_p})$ from the {\bf TPC dE/dx} measurement. The probabilities for
  the kaon and proton hypothesis are taken from the RPRODE routine. If no TPC information
  is available the input is set to 0.
\item the {\bf kaon probability} from the {\bf TPC dE/dx} given by the RPRODE routine.
\item The probability of the {\bf kaon pull} from the TPC. If no TPC information is available
  the input is set to 0.
\item the logarithm of the {\bf ratio of the dE/dx probabilities} of the {\bf kaon}
  to the {\bf pion} hypothesis for kaon separation from pions with the {\bf vertex detector}.
  For this purpose the dE/dx information for the different hypotheses have been fitted over the
  whole momentum region covered by the vertex detector. For a given measurement and hypothesis
  the distance of the dE/dx signal from the fitted function is divided by the error on the
  measurement to obtain the deviation from this hypothesis in standard deviations.
\item the logarithm of the {\bf ratio of the dE/dx probabilities} of the {\bf kaon}
  to the {\bf proton} hypothesis for kaon separation from protons with the {\bf vertex detector}.
\item the {\bf probability of the kaon VD dE/dx} hypothesis.
\item the {\bf number of TPC wires} hit by the particle
\item the {\bf momentum} of the particle
\end{enumerate}
For the proton net the same net topology was used as for the kaons. For the training
$\approx$2,000 protons were used mixed with the same amount of background. The kaon inputs
of the vertex detector were changed to proton variables.
%
\section{Performance}
%
%
\subsection{Kaon net}
%
%
%
\begin{figure}[t]
\begin{minipage}[t]{0.495\textwidth}
\begin{flushleft}
\leavevmode 
\includegraphics[bb=16 145 560 701, width=0.96\textwidth]
                {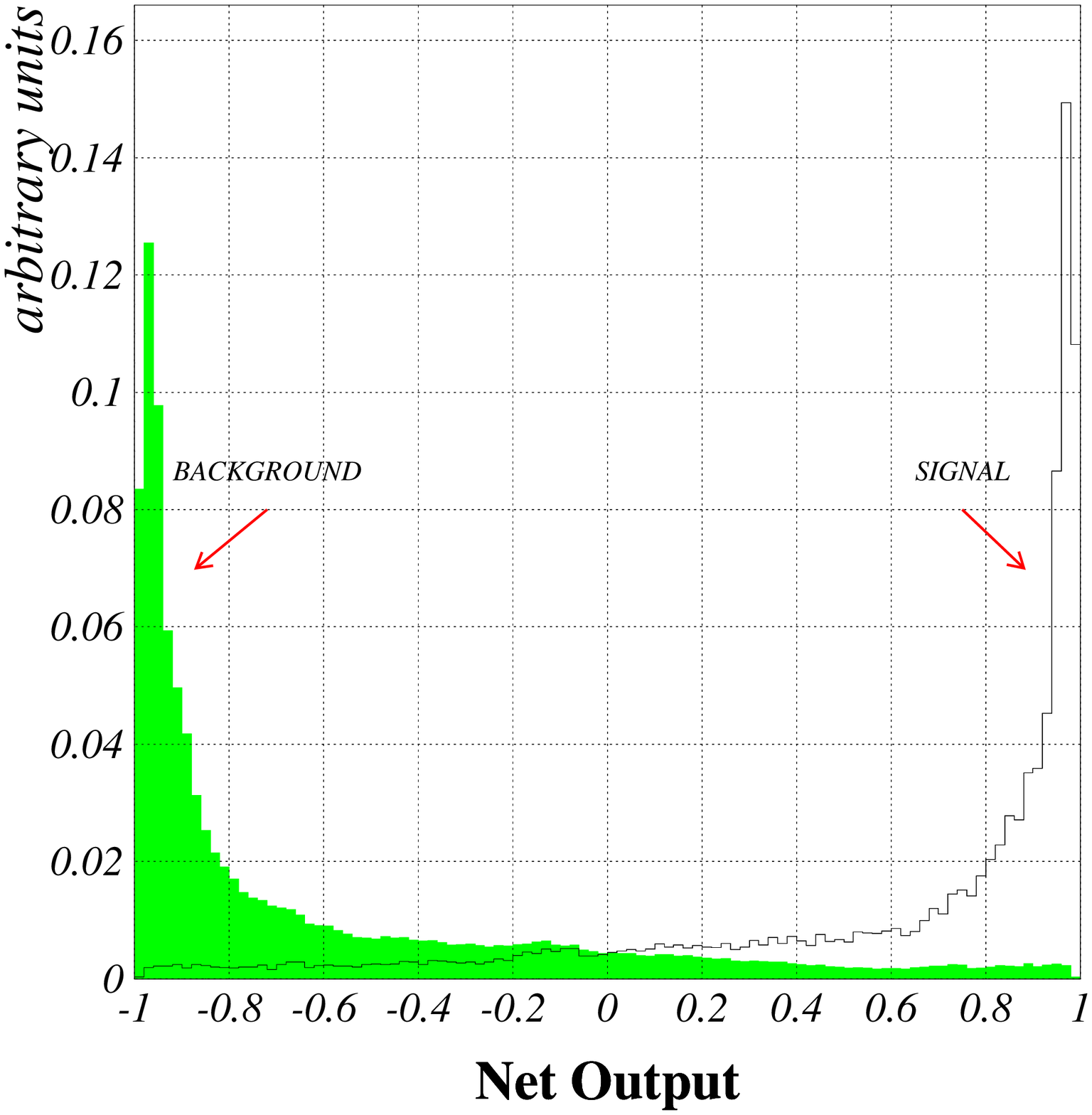}
\end{flushleft}
\end{minipage}
\hfill
\begin{minipage}[t]{0.495\textwidth}
\begin{flushright}
\leavevmode 
\includegraphics[bb=18 145 560 680, width=0.96\textwidth]
                {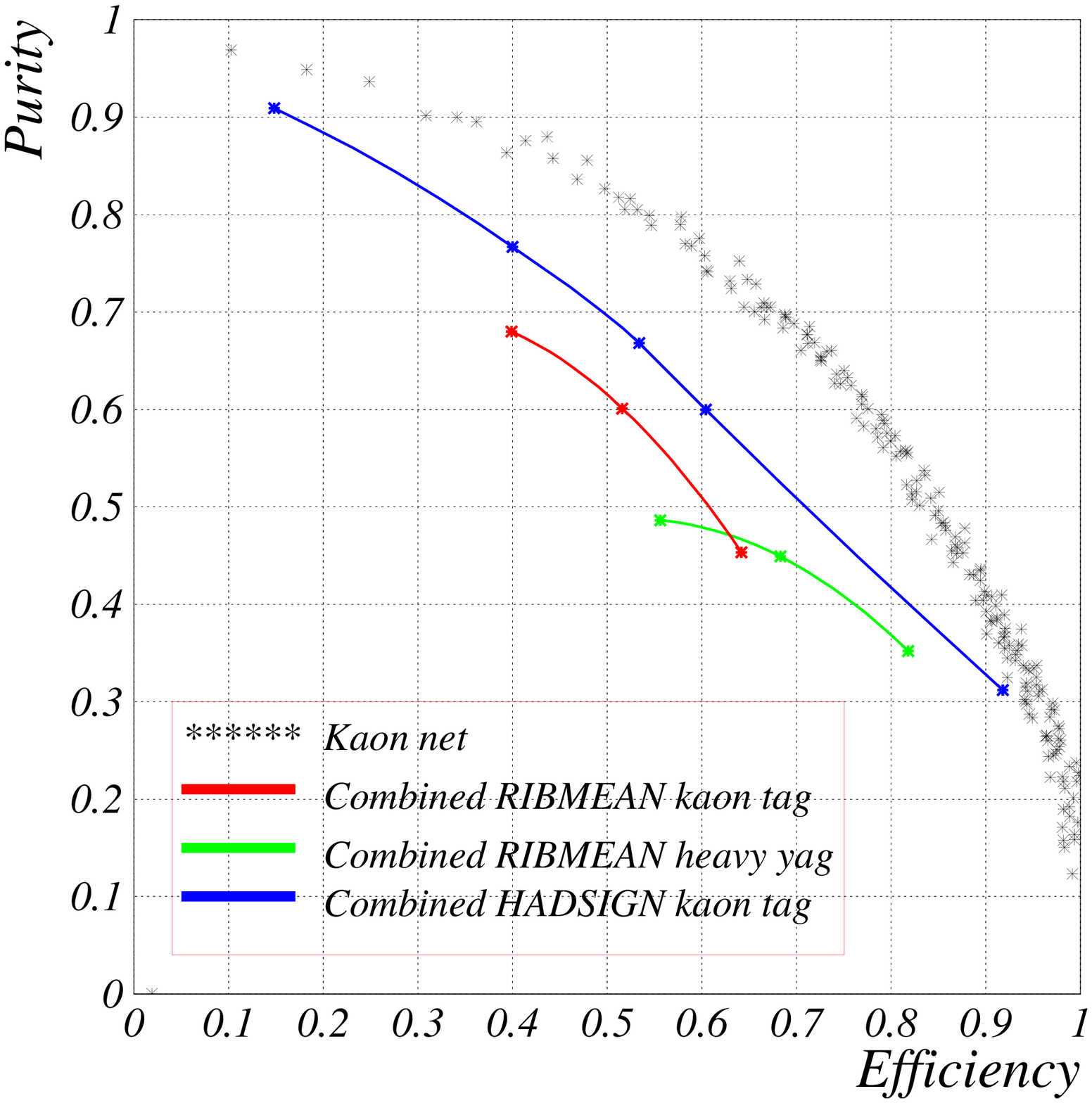}
\end{flushright}
\end{minipage}
\caption[Performance of MACRIB for kaons.]
         {\label{fig:kanet} Left: Signal-Background separation of MACRIB for kaons with
           full RICH information.
           Signal is accumulated at +1 and background at -1. The separation of the two
           classes is well done. This allows an efficient separation of kaons from the
           background, mainly pions. Right: The momentum averaged efficiency purity
           plot of MACRIB for kaons compared to the combined (RICH gas and liquid +
           dE/dx) information from HADSIGN and RIBMEAN.}
\end{figure}
For the separation of background and signal the target of the net outputs were
set to -1 and 1 respectively. The output of the net with full RICH information can be seen
on the left side of figure \ref{fig:kanet}. A clear separation between kaons and background can be
observed. Compared with the combined HADSIGN kaon tag,
the combined RIBMEAN kaon tag and RIBMEAN heavy tag one can see  that the kaon net of MACRIB has a far
better efficiency at the same purity (right side of fig. \ref{fig:kanet}).

\begin{figure}[t]
\begin{minipage}[t]{0.495\textwidth}
\begin{flushleft}
\leavevmode 
\includegraphics[bb=16 145 560 701, width=0.96\textwidth]
                {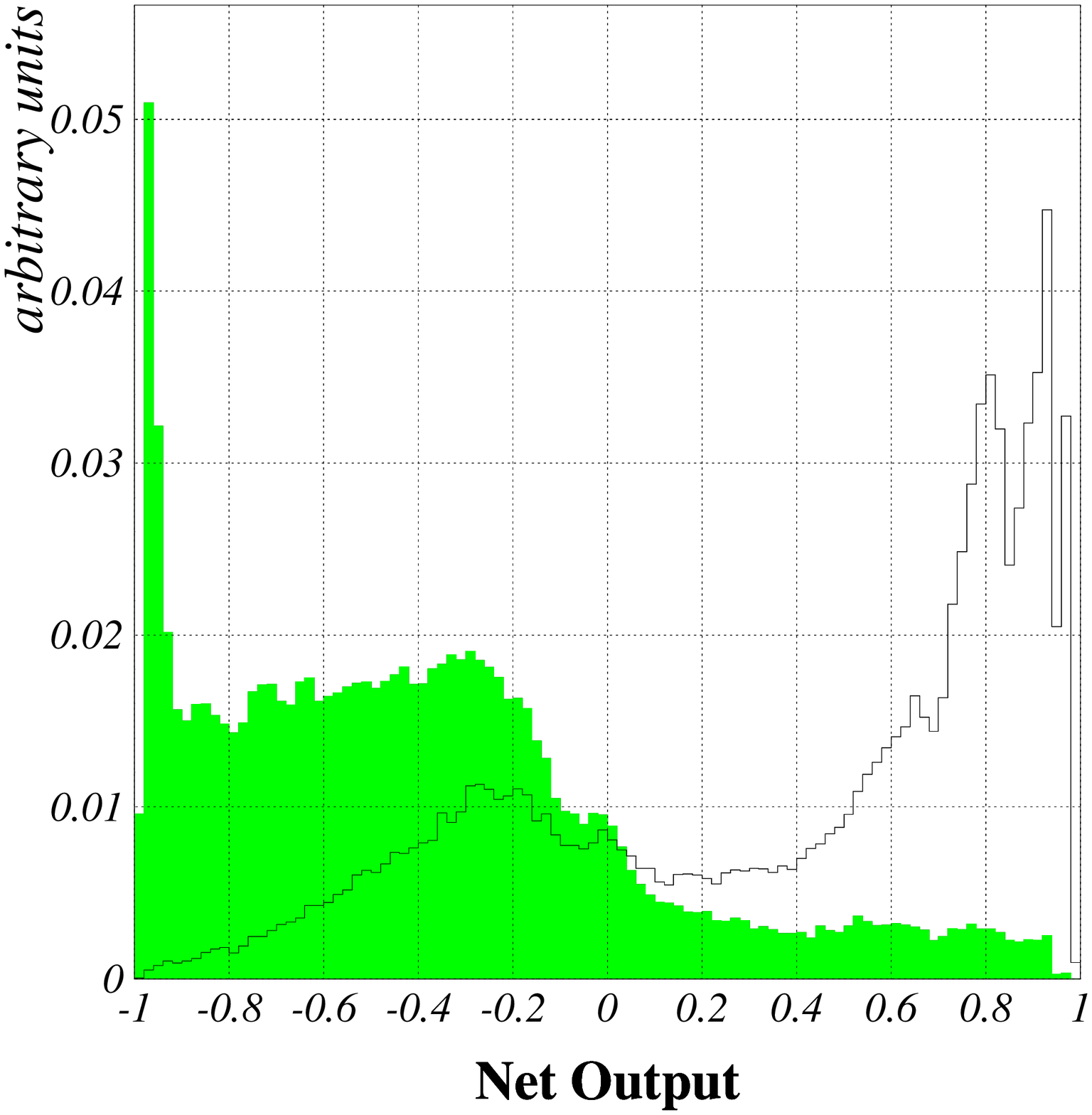}
\end{flushleft}
\end{minipage}
\hfill
\begin{minipage}[t]{0.495\textwidth}
\begin{flushright}
\leavevmode 
\includegraphics[bb=18 145 560 680, width=0.96\textwidth]
                {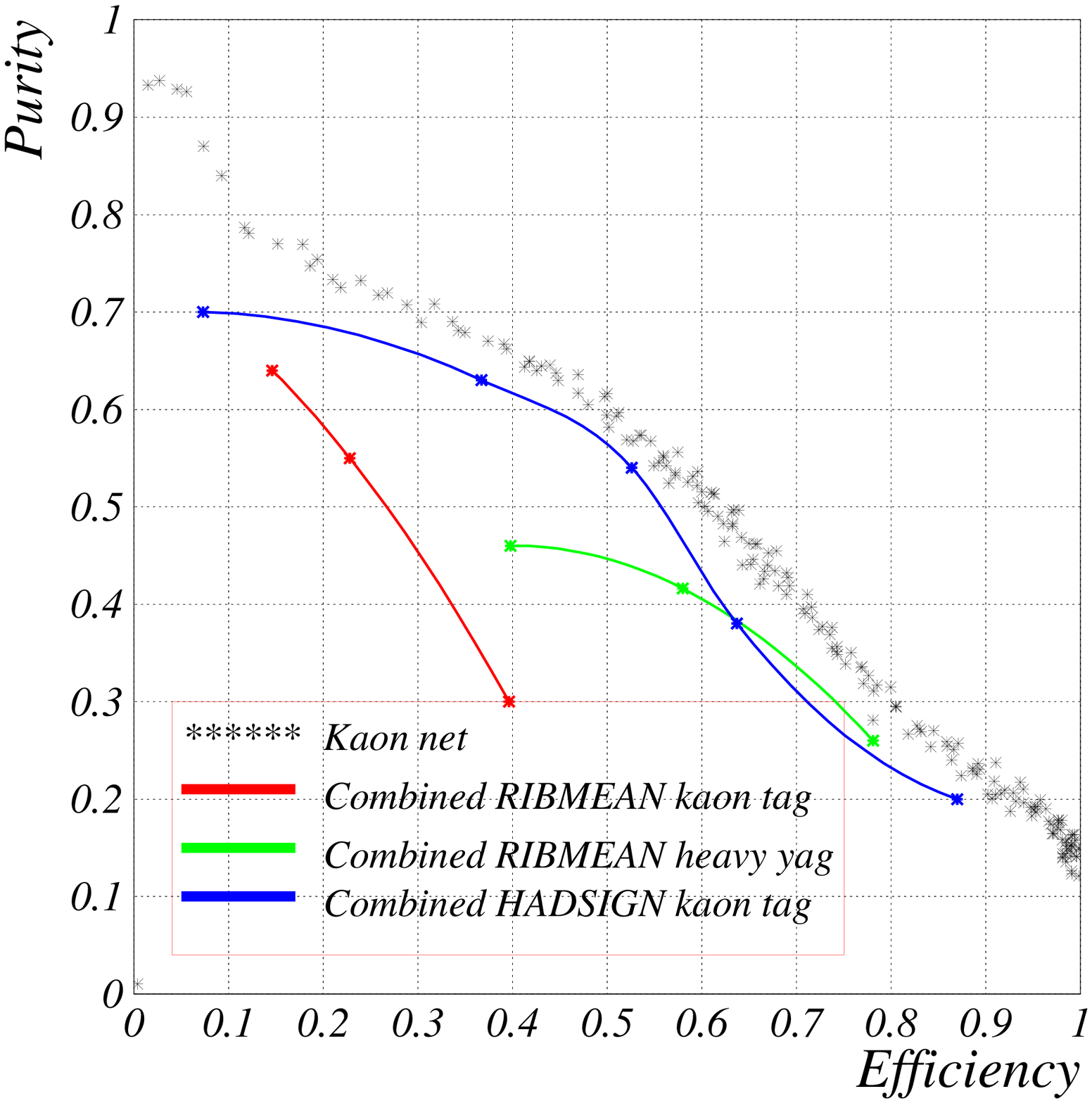}
\end{flushright}
\end{minipage}
\caption[Performance of MACRIB for kaons.]
         {\label{fig:kanetnl} Left: Signal-Background separation of MACRIB for kaons
           without liquid RICH information. The separation of the two
           classes can not be done as efficiently as with full RICH information but
           there is still an improvement compared to the combined RIBMEAN and HADSIGN
           tags.}
\end{figure}

As one would expect, the performance of the network without liquid RICH information can not
be as good as with the working liquid radiator. The missing information results in a class
of particles that lie in the momentum range that is not covered by any detector. These get a net output
that lies somewhere between -1 and 1. The bump in the middle of the
plot at the left side of figure \ref{fig:kanetnl} comes from this class. The comparison
with the HADSIGN and RIBMEAN tags shows again a significant improvement. The fact
that one can cut on a continuous variable instead of taking discrete tags makes it in addition a more
flexible tool.
%
%
%
%
\begin{figure}[bt]
\begin{minipage}[t]{0.495\textwidth}
\begin{flushleft}
\leavevmode 
\includegraphics[bb=16 145 560 701, width=0.96\textwidth]
                {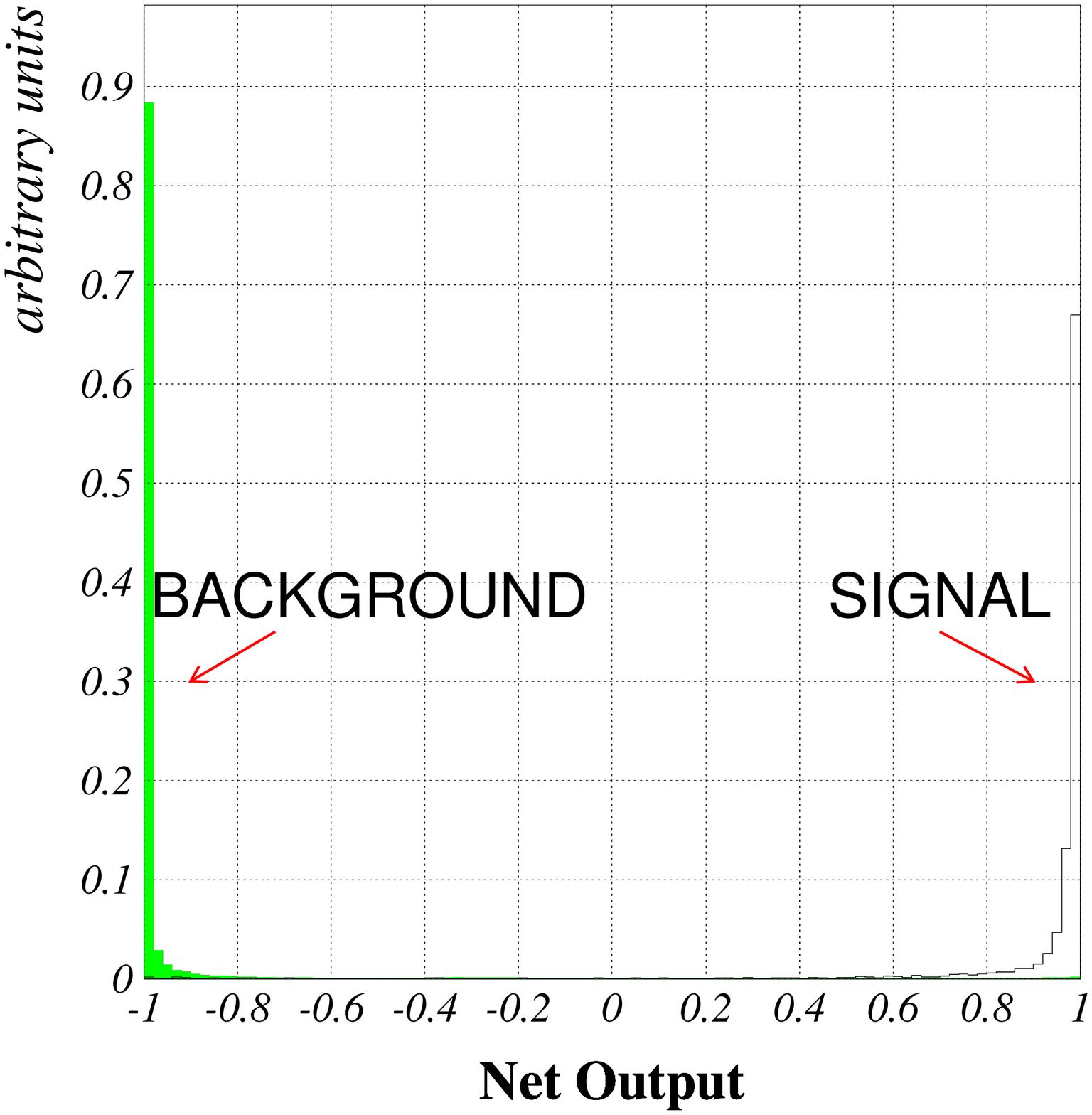}
\end{flushleft}
\end{minipage}
\hfill
\begin{minipage}[t]{0.495\textwidth}
\begin{flushright}
\leavevmode 
\includegraphics[bb=18 145 560 680, width=0.96\textwidth]
                {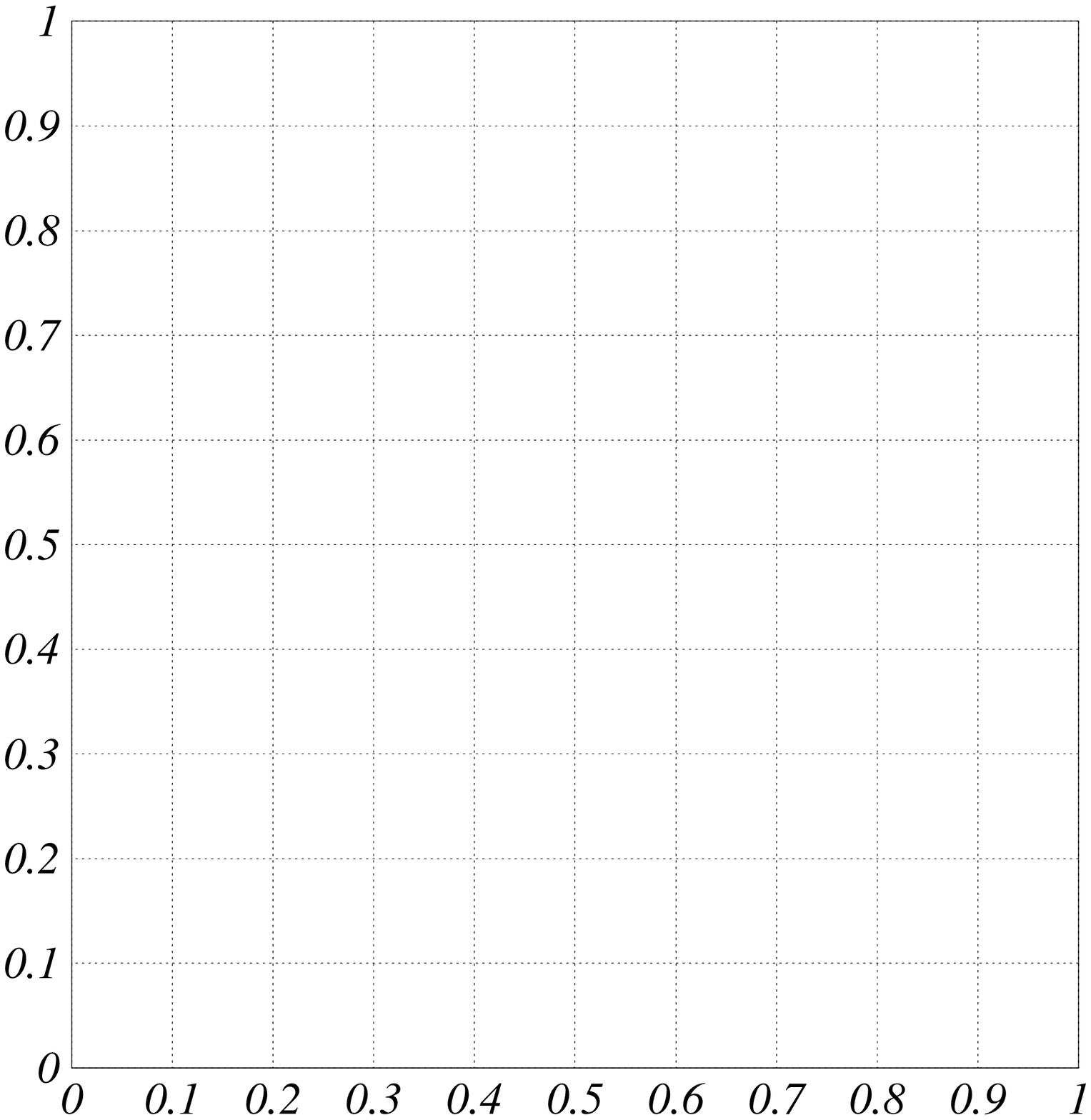}
\end{flushright}
\end{minipage}
\caption[Performance of MACRIB for kaons with low momentum.]
         {\label{fig:kanetlm} Left: The output of the MACRIB kaon net for low momentum kaons. Both
           TPC and VD informations have been used as input for the net. Right: The large gain in
           efficiency compared to RIBMEAN and HADSIGN comes largely from the independent
           dE/dx information of the vertex detector.}
\end{figure}
\begin{figure}[p]
  \centering
  \epsfig{file=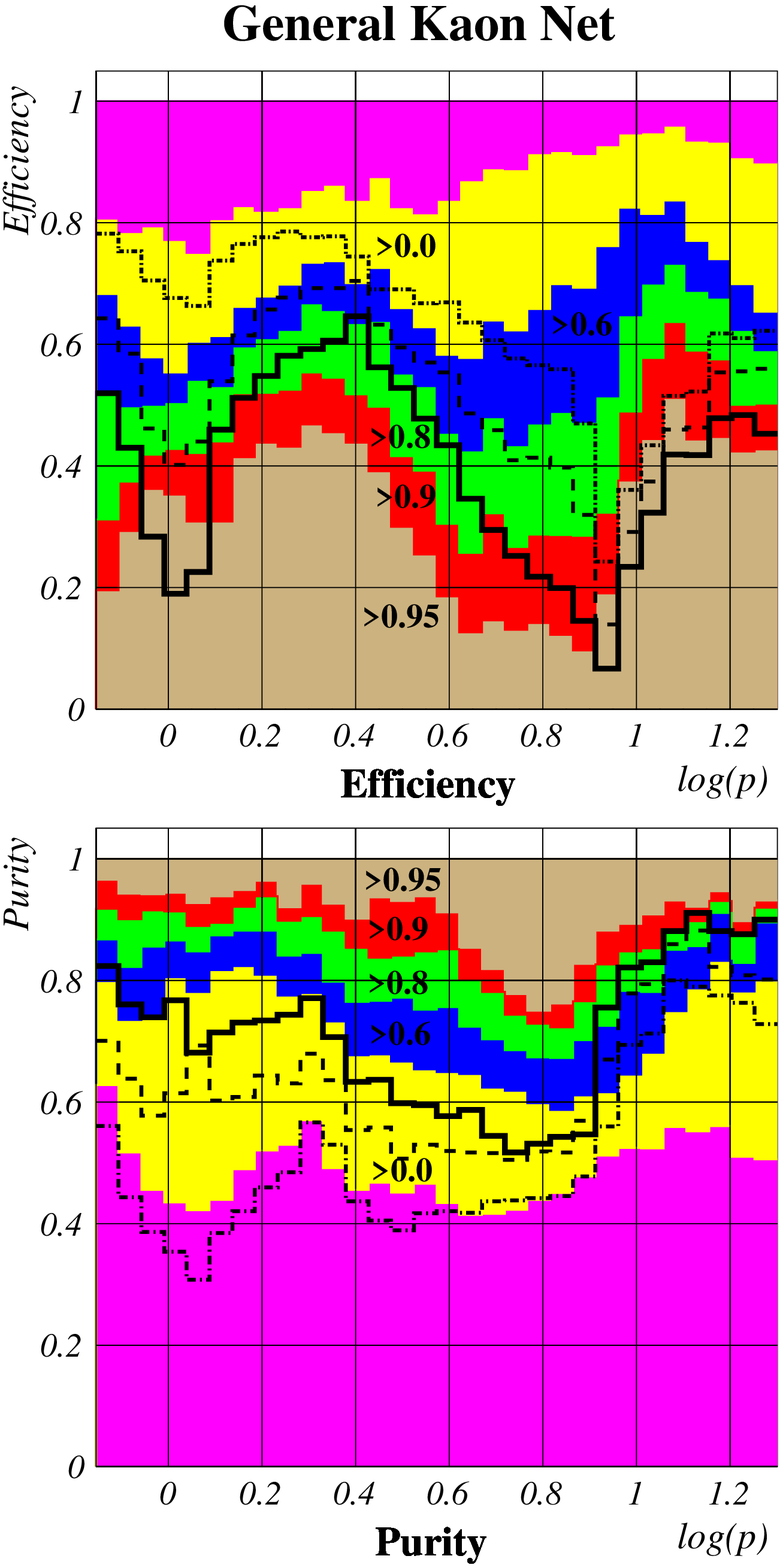,width=10cm}
  \caption[Efficiency and purity of MACRIB vs. momentum for kaons] {Efficiency and
           purity of MACRIB vs. momentum for kaons
           compared to the RIBMEAN performance. The solid line corresponds
           to the tight kaon tag, the dashed line to the standard kaon
           tag and the dash-dotted line to the loose tag.}
  \label{fig:epomknet}
\end{figure}

The output of the kaon net below 0.7~GeV is displayed at the left side of figure \ref{fig:kanetlm}.
Since in the TPC the kaon band is well separated from the pion band (see figure \ref{fig:pid}) a very
pure signal to background ratio can be achieved by only considering TPC variables. However, for
cases where no TPC measurement is available, HADSIGN and RIBMEAN can provide no tag. This happens,
when less then 30 wires are hit or if the kaon pull is larger then two sigmas. By combining the
TPC measurement with the information coming from
the vertex detector, a very good kaon identification can be achieved even for these particles.
This further
information is one of the reasons for the large gain in efficiency at same purity compared to the
HADSIGN and RIBMEAN tags shown at the right side of figure \ref{fig:kanetlm}.

Since purity and efficiency depend on momentum the efficiency purity plots in the figures
\ref{fig:kanet},\ref{fig:kanetnl} and \ref{fig:kanetlm} only show the
averaged purity and the averaged efficiency in hadronic Z decays. Figure~\ref{fig:epomknet} shows the
performance of MACRIB over the whole momentum spectrum. 
The three levels of the combined RIBMEAN kaon tag are compared to different cuts on the kaon net
output represented by different colours. The colour codes are the same in both plots for the
efficiency and purity. Note that especially the efficiency hole of the RIBMEAN tags below 8 GeV is
filled up by MACRIB.
%
%
%
\begin{figure}[t]
\begin{minipage}[t]{0.495\textwidth}
\begin{flushleft}
\leavevmode 
\includegraphics[bb=16 145 560 701, width=0.9\textwidth]
                {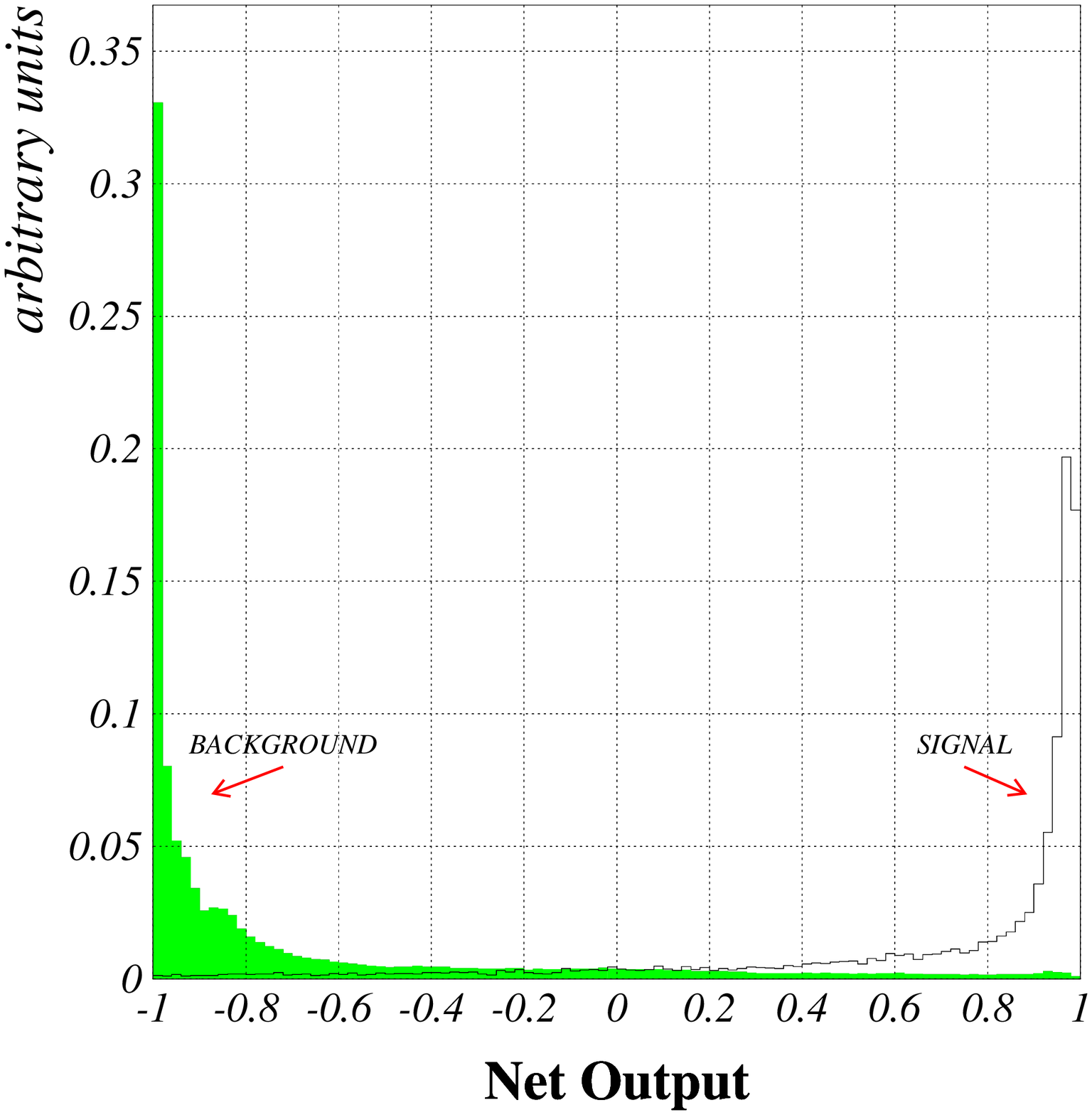}
\end{flushleft}
\end{minipage}
\hfill
\begin{minipage}[t]{0.495\textwidth}
\begin{flushright}
\leavevmode 
\includegraphics[bb=18 145 560 680, width=0.9\textwidth]
                {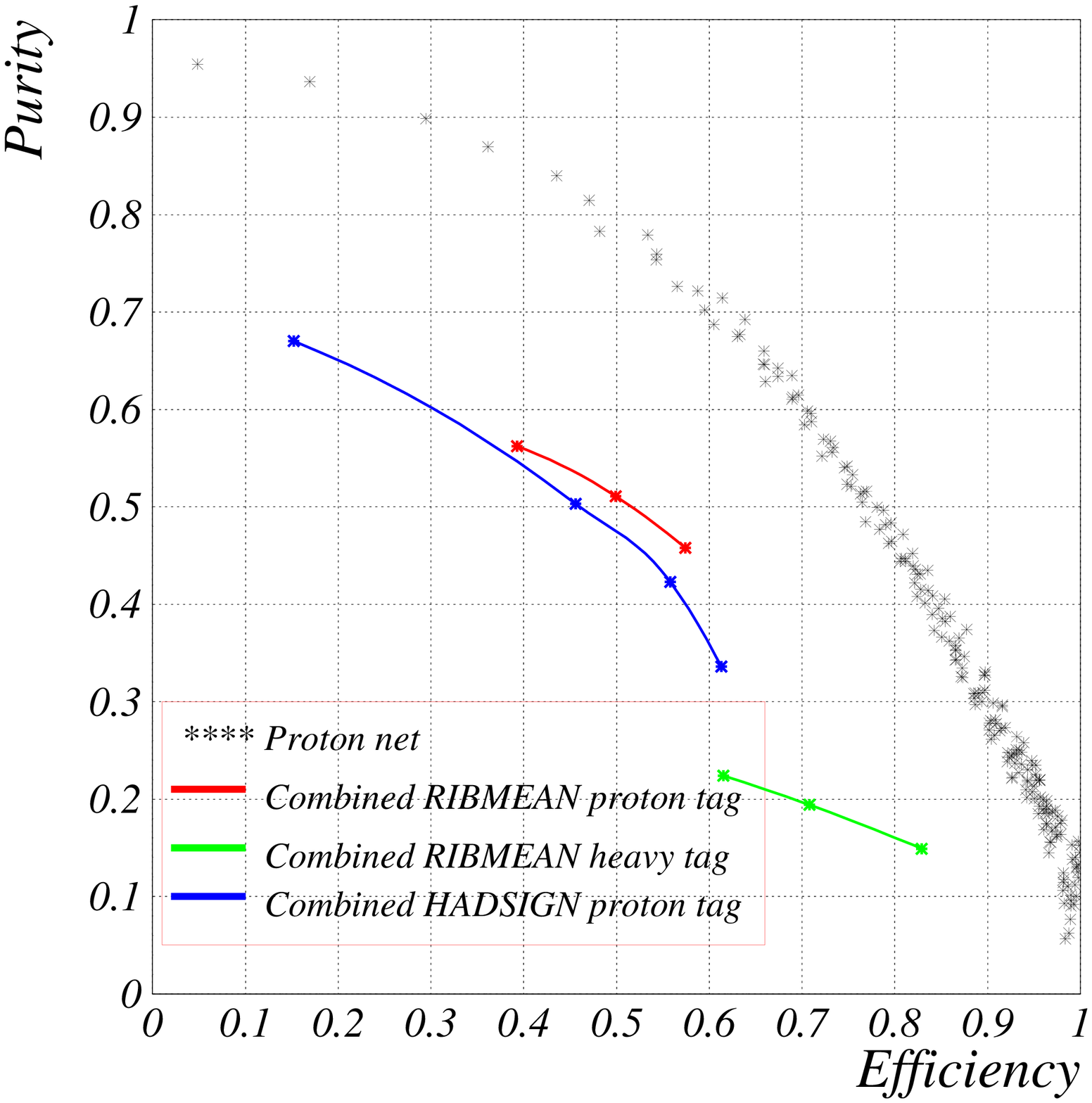}
\end{flushright}
\end{minipage}
\caption[Proton net output with full RICH information.]
         {\label{fig:prnet} Left: The separation of background and signal of MACRIB for protons.
           Like for kaons, both classes are well pulled to a peak at -1. and +1. respectively.
           Right: The purity vs. efficiency of the MACRIB package for protons
           compared to the performance of the RIBMEAN and HADSIGN tags.}
\end{figure}
\subsection{Proton net}
The output of the net with liquid RICH information for protons and background can be
seen in \ref{fig:prnet} together with the comparison of the purity
over efficiency plot of the MACRIB routine with that of the combined tags of
RIBMEAN and HADSIGN. At high purity a large gain in efficiency is achieved.

%
%
\begin{figure}[t]
\begin{minipage}[t]{0.495\textwidth}
\begin{flushleft}
\leavevmode 
\includegraphics[bb=16 145 560 701, width=0.9\textwidth]
                {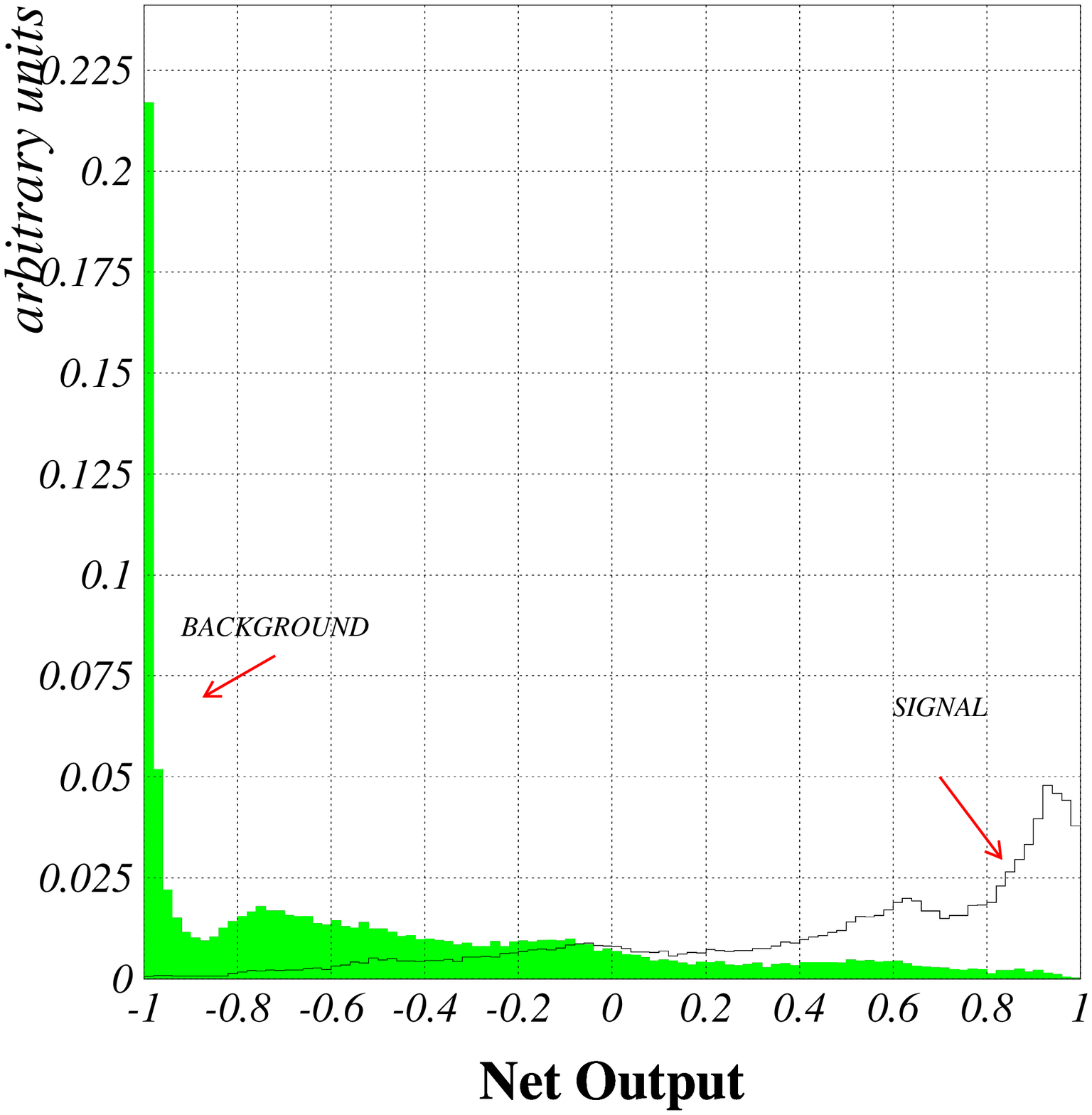}
\end{flushleft}
\end{minipage}
\hfill
\begin{minipage}[t]{0.495\textwidth}
\begin{flushright}
\leavevmode 
\includegraphics[bb=18 145 560 680, width=0.9\textwidth]
                {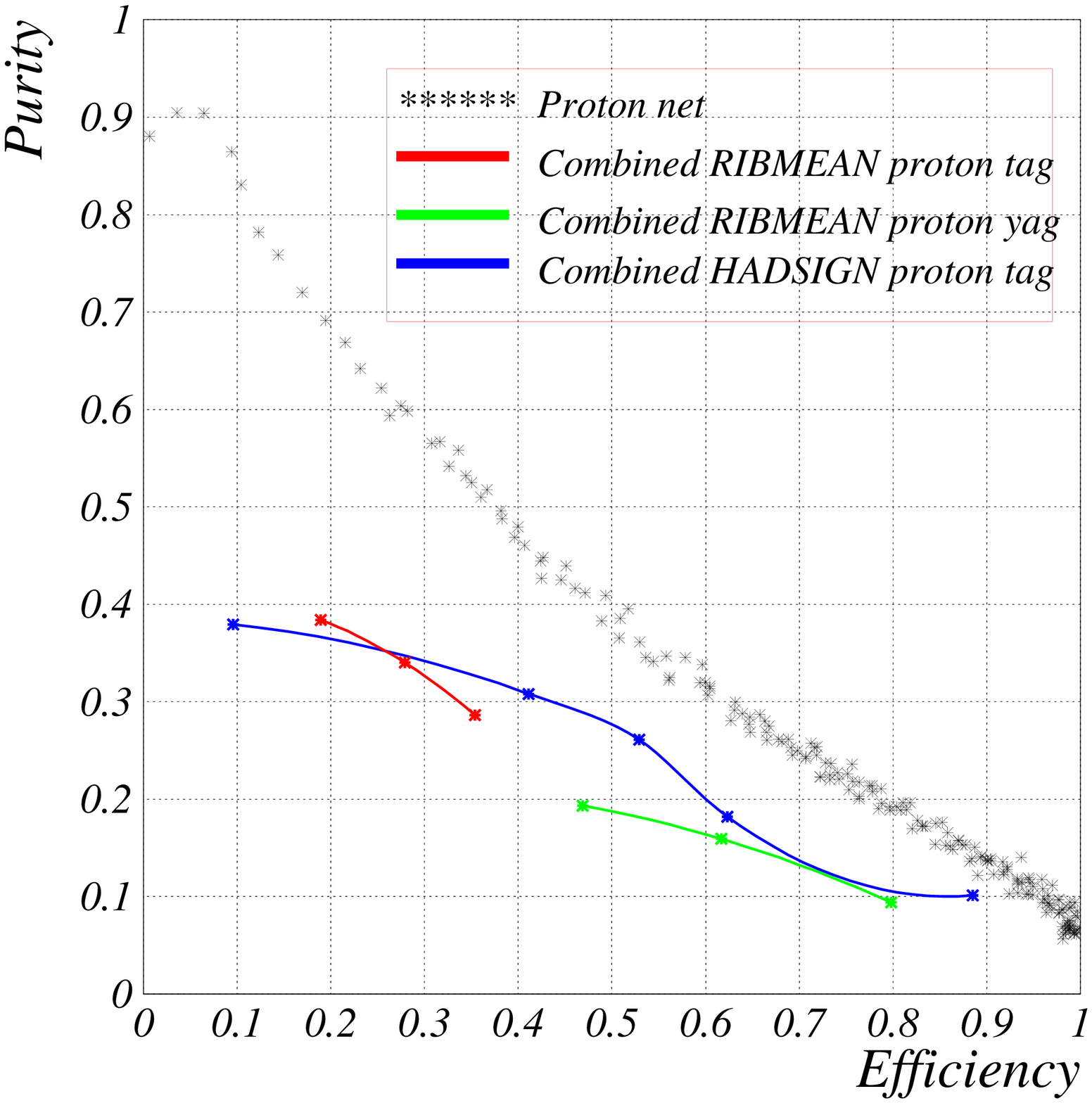}
\end{flushright}
\end{minipage}
\caption[Signal/background separation and purity vs. efficiency for protons.]
         {\label{fig:prnetnl} Left: The separation of background and signal of MACRIB for protons.
           Like for kaons, both classes are well pulled to a peak at -1. and +1. respectively.
           Right: The purity vs. efficiency of the MACRIB package for protons
           compared to the performance of the RIBMEAN and HADSIGN tags.}
\end{figure}
%
%
\begin{figure}[bt]
\begin{minipage}[t]{0.495\textwidth}
\begin{flushleft}
\leavevmode 
\includegraphics[bb=16 145 560 701, width=0.96\textwidth]
                {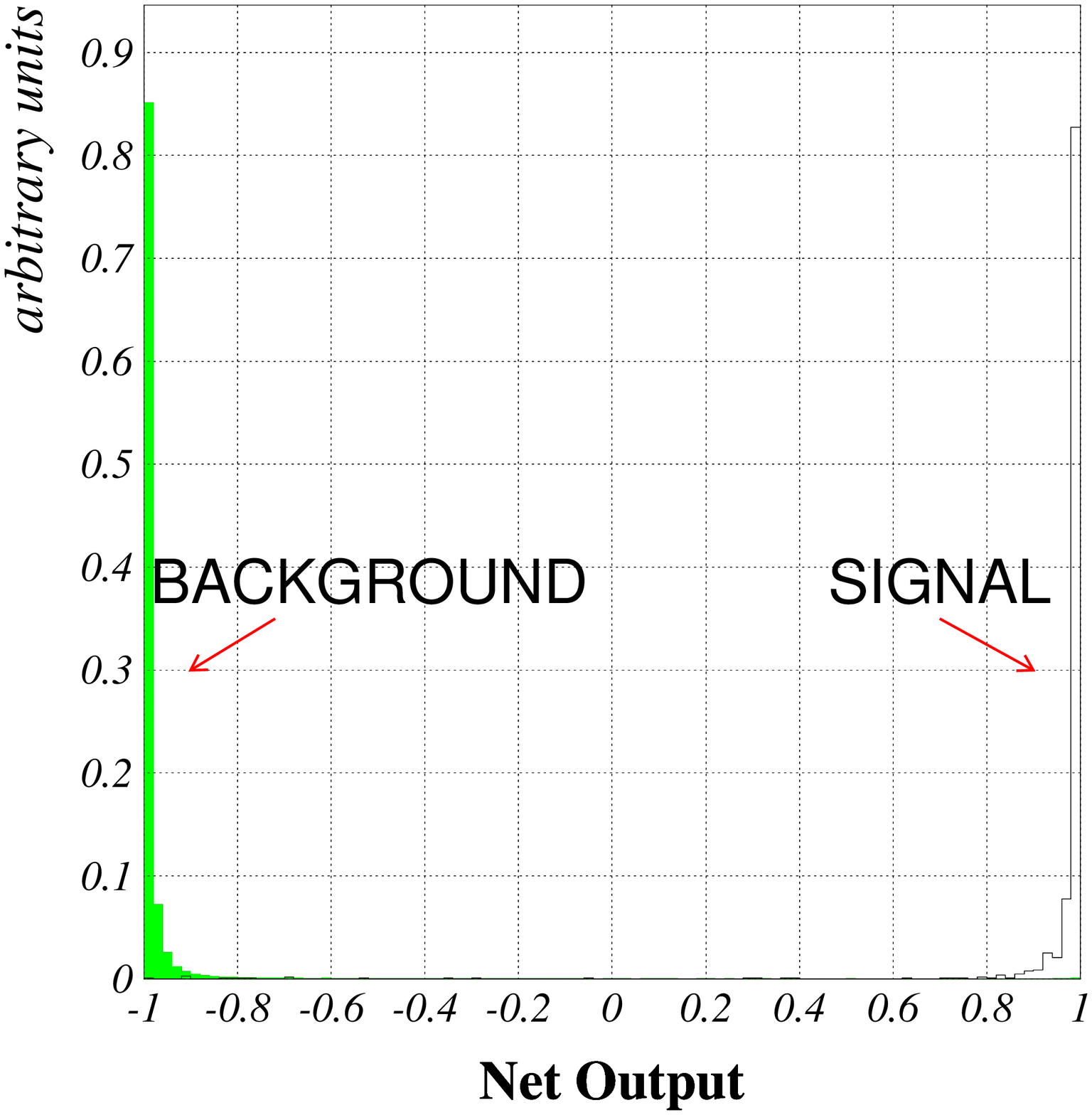}
\end{flushleft}
\end{minipage}
\hfill
\begin{minipage}[t]{0.495\textwidth}
\begin{flushright}
\leavevmode 
\includegraphics[bb=18 145 560 680, width=0.96\textwidth]
                {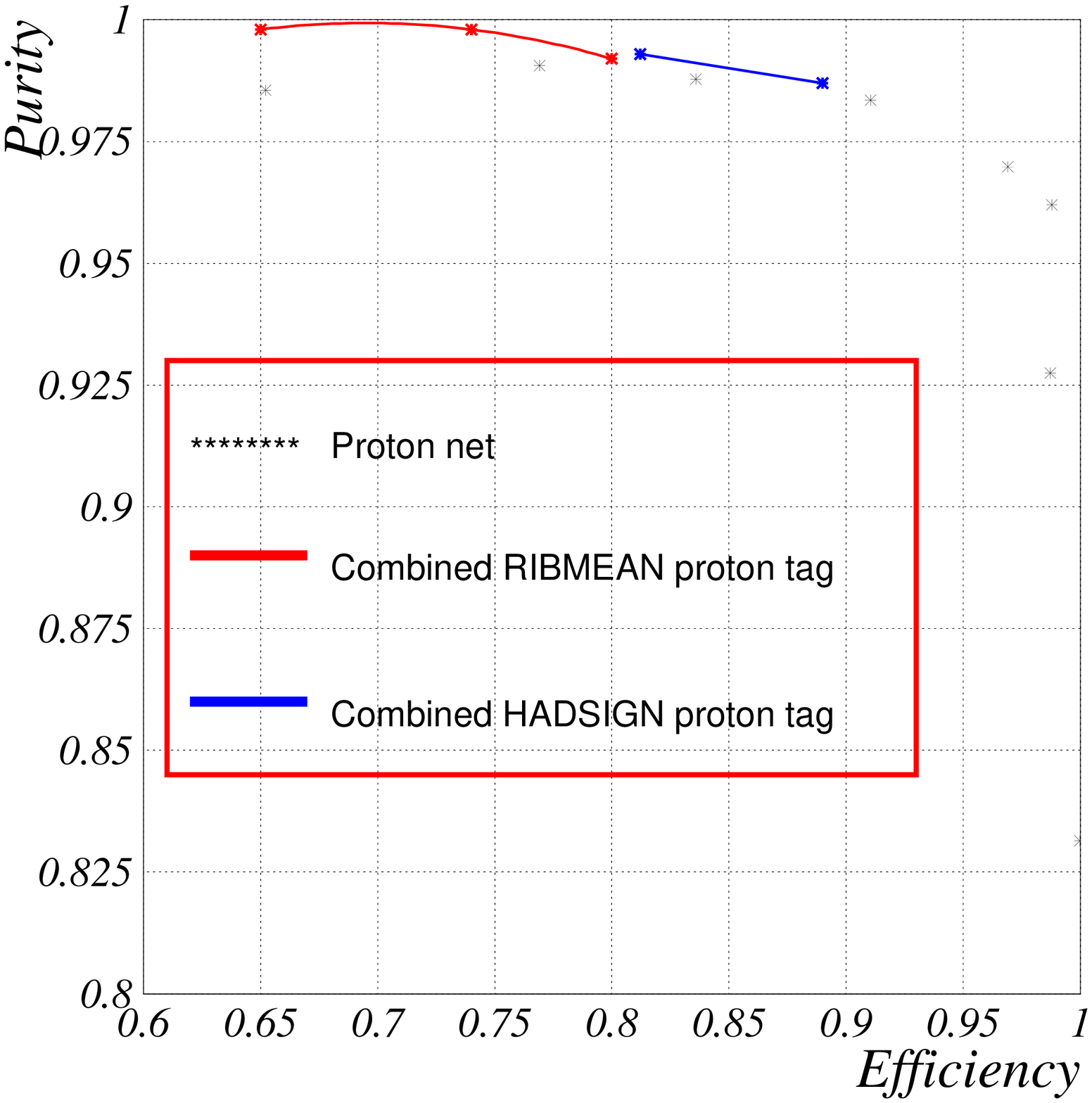}
\end{flushright}
\end{minipage}
\caption[Performance of MACRIB for protons with low momentum.]
         {\label{fig:prnetlm} Left: Signal and background are accumulated at the extreme ends of the
           net output plot. Right: Due to the very good separability of protons from background a
           very high level of purity and efficiency can be achieved. Note that only a small window
           of the efficiency purity plot is shown. MACRIB has a better efficiency than the RIBMEAN
           and HADSIGN tags, due to the additional VD information.}
\end{figure}
\begin{figure}[p]
  \centering
  \epsfig{file=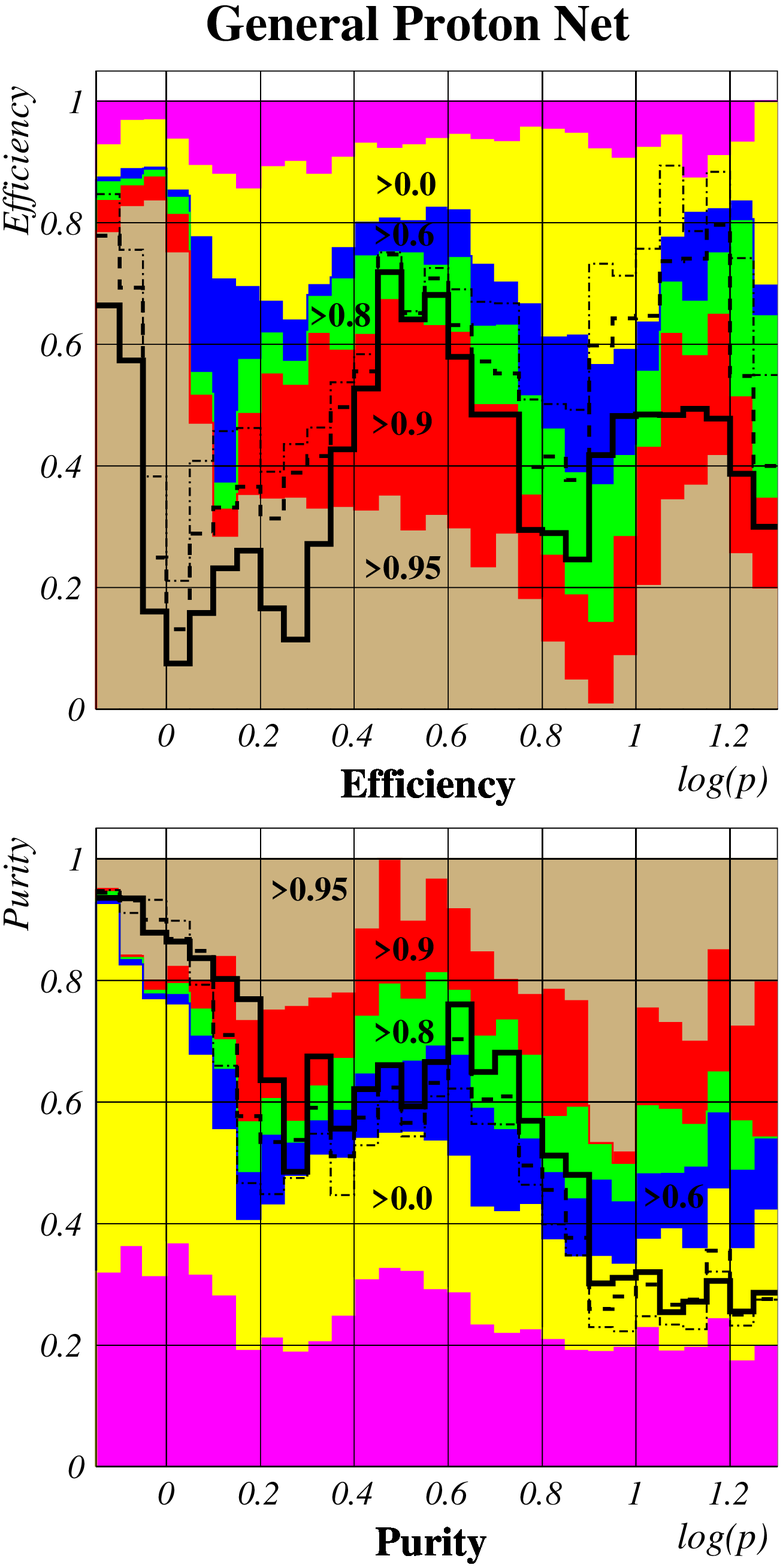,width=10cm}
  \caption[Efficiency and purity of MACRIB vs. momentum for protons]
          {Efficiency and purity  of MACRIB vs. momentum for protons
           compared to the RIBMEAN performance. The solid line corresponds
           to the tight proton tag, the dashed line to the standard proton
           tag and the dash-dotted line to the loose tag.}
  \label{fig:epprotmom}
\end{figure}
As for the kaons, the performance of the net goes down if no liquid RICH information is available.
Though, the proton net deals better with the loss of this information than the
kaon net. This lies in the fact that the VD dE/dx bands of protons and pions are still separated in
the momentum region usually covered by the liquid radiator. The bump, representing the class of particles
with no detector information or with ambiguous information, in the middle of the performance plot
at the left side of figure \ref{fig:prnetnl} can be found here too, but  at a lower level  than for the kaon net.
The efficiency purity plot shows the possibility of going to very pure proton samples. This
option was not available with the RIBMEAN and HADSIGN tags, as can be seen at the right side of figure
\ref{fig:prnetnl}.

For particles with momenta below 0.7~GeV the net output is shown in figure \ref{fig:prnetlm} (left). The
proton and pion bands are separated by many sigmas in both the TPC and vertex detectors. This makes
a very pure and efficient identification possible. The comparison between MACRIB, RIBMEAN and HADSIGN
can be seen at the right side of the same figure. The particles having no RIBMEAN or HADSIGN tags
contribute to the efficiency gain in MACRIB, by utilising the independent dE/dx measurement provided by the
VD.

The momentum dependence of the efficiency and purity can be seen in figure \ref{fig:epprotmom}
together with the
loose, standard and tight tags of the RIBMEAN routine.
%
\subsection{Simulation-Data Agreement}
%
\begin{figure}[t]
\begin{minipage}[t]{0.495\textwidth}
\begin{flushleft}
\leavevmode 
\includegraphics[bb=65 185 490 665, width=0.96\textwidth, clip=]{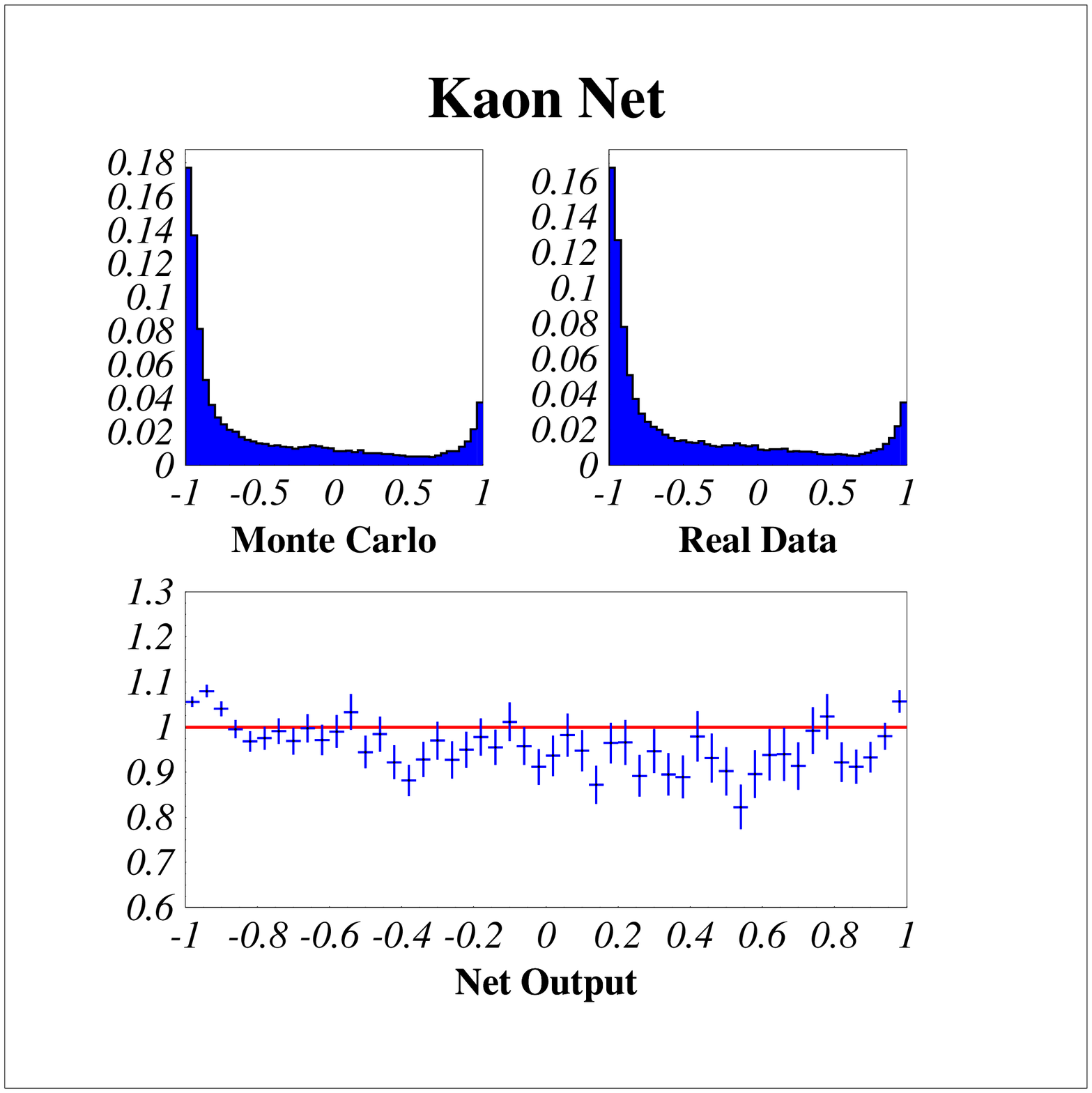}
\end{flushleft}
\end{minipage}
\hfill
\begin{minipage}[t]{0.495\textwidth}
\begin{flushright}
\leavevmode 
\includegraphics[bb=65 185 490 665, width=0.96\textwidth, clip=]{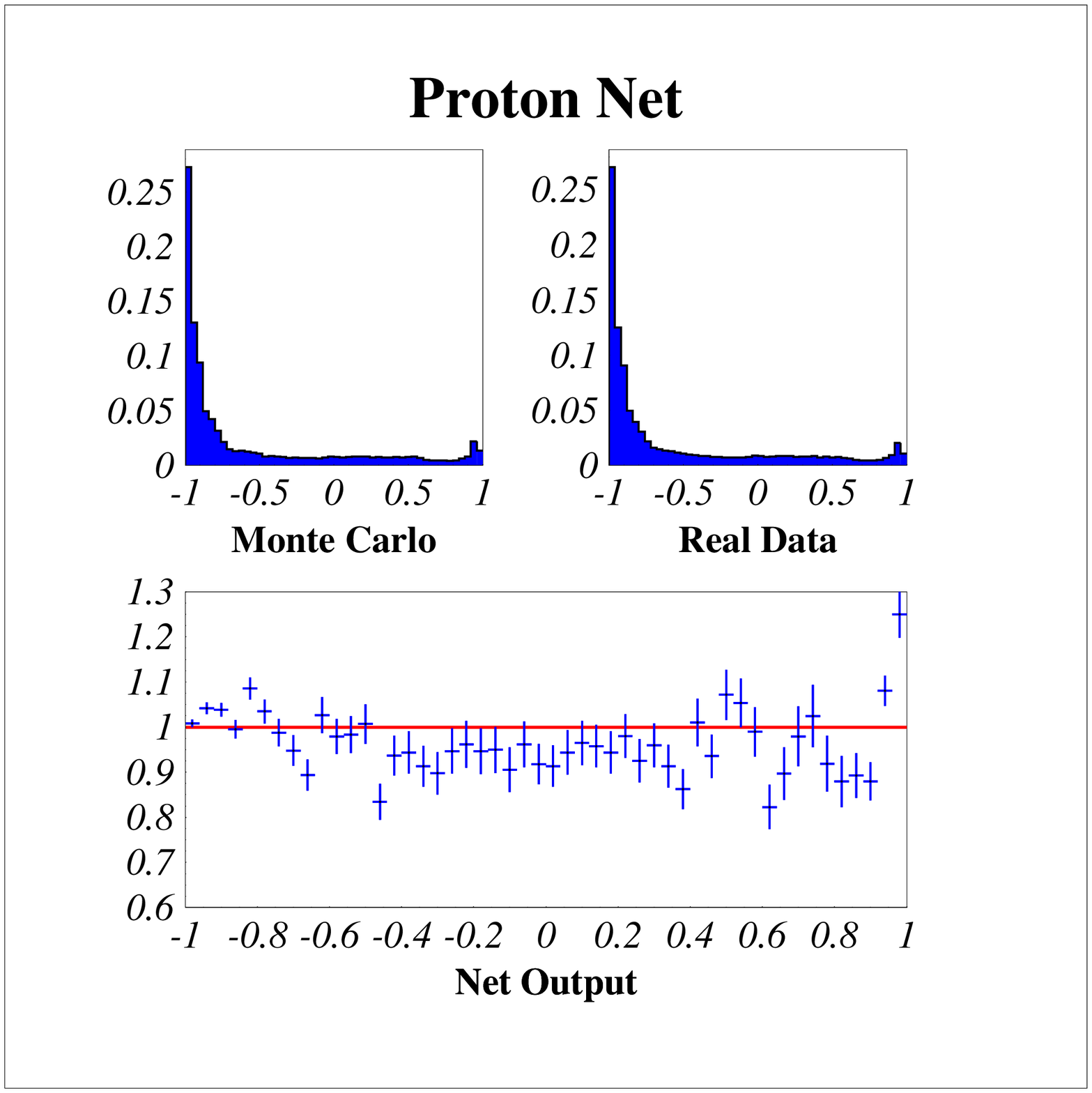}
\end{flushright}
\end{minipage}
  \caption[The Monte-Carlo-Data Agreement of the MACRIB output.]
          {Left: The output of MACRIB for kaons for an independent MC sample and
           real data. Below the ratio of the normalised distributions can be
           seen (MC divided by the real data). Right: The output of MACRIB for protons, with the ratio
           of the normalised distributions.}
  \label{fig:outcomp}
\end{figure}
For the test of the Simulation-Data agreement of the network output we used an independent
Monte Carlo sample, which has not been used in the training of the nets.  For
the case of fully working RICH detectors the comparison between real data
and Monte Carlo is shown in figures \ref{fig:outcomp}. The distributions have been normalised and the
simulated sample divided by the real data. The agreement is reasonable over most of the network output range.
At the far ends of the plot a surplus of the MC can be
observed. This is due to the more ideal conditions of the simulation, so that the net
can assign an output value closer to the edges.
%
\section{MACRIB - How to get the net output}
%
We have written a simple stand-alone-routine MACRIB returning the kaon- and
the proton-net outputs for a given track with the PA-pointer LPA. This routine
is included in the BSAURUS package, since it uses the network routines of
BSAURUS. The user has to include the BSAURUS library by including BSAURUS
into the argument list of dellib
in the link step of the job in order to have access to the routine. \\
Before calling MACRIB for a track one first has to call once per event\cite{RICH}:
\begin{itemize}
  \item CALL RICHID
  \item CALL RIDECR
  \item CALL GETMINE(MYTYPE) \\
    where MYTYPE=1 means LPA/tanagra, MYTYPE=2 means VECSUB
\end{itemize}
After that MACRIB can be called with the input argument LPA:
\begin{center}
  CALL MACRIB(LPA,XKAONNET,XPROTONNET)
\end{center}
BSAURUS users can access the net outputs via the BSAURUS commons
\begin{center}
  BSPAR(IBP\_MACK,IPART)
  BSPAR(IBP\_MACP,IPART)
\end{center}
on the track basis.
The output variables are the kaon net output XKAONNET and the proton
net output XPROTONNET. They are real variables with values between -1. and 1.
XKAONNET/XPROTONNET $\to$ -1. means identification as not a kaon/proton
whereas XKAONNET/XPROTONNET $\to$ 1. means identified as kaon/proton.
Tracks that fail to pass the cuts in MACRIB get the value -2.
%
\section{Sample plots}
%
\begin{figure}[t]
\begin{minipage}[t]{0.495\textwidth}
\begin{flushleft}
\leavevmode 
\includegraphics[bb= 100 142 490 690, width=0.94\textwidth]{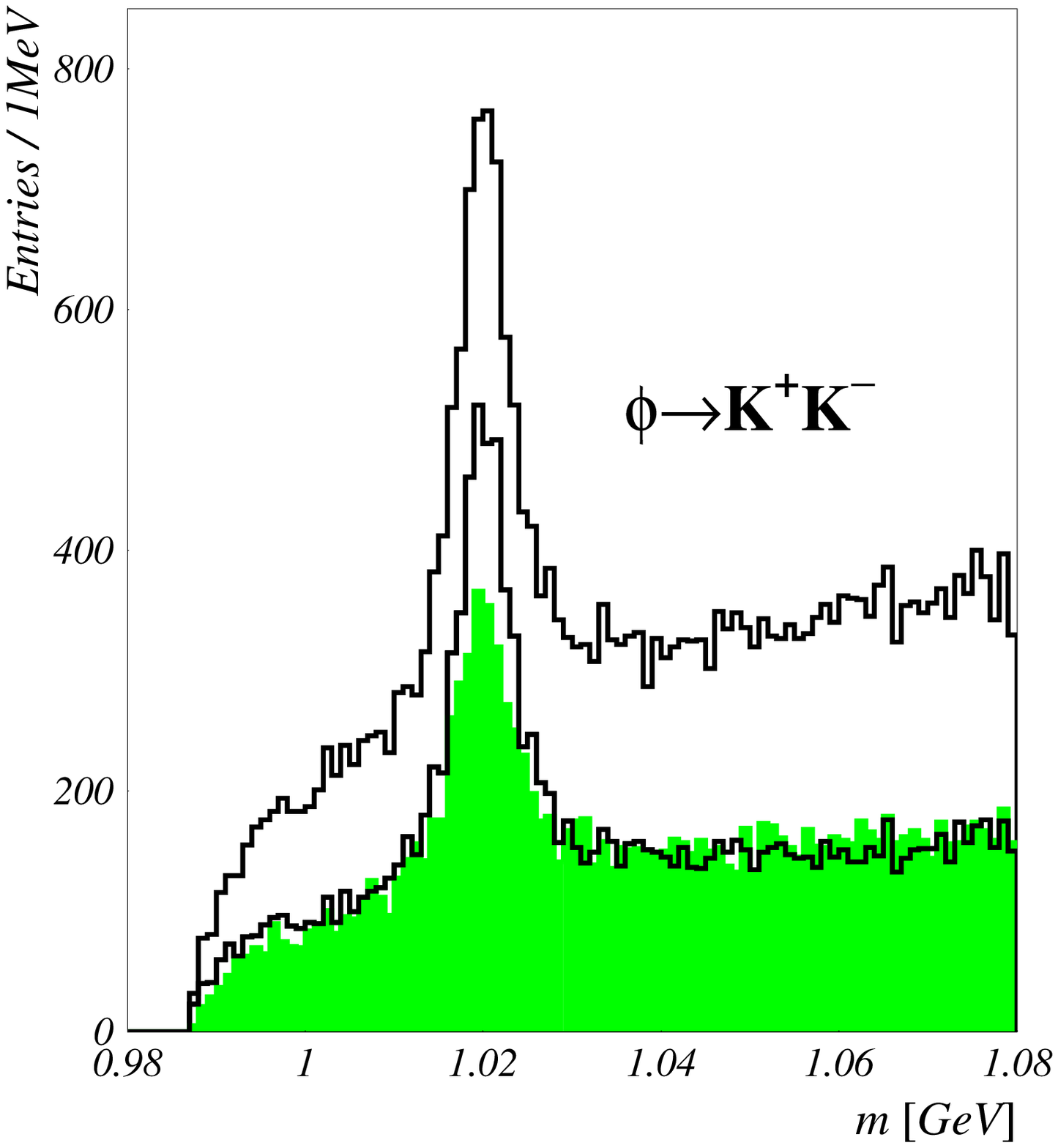}
\end{flushleft}
\end{minipage}
\hfill
\begin{minipage}[t]{0.495\textwidth}
\begin{flushright}
\leavevmode 
\includegraphics[bb= 100 142 490 690, width=0.94\textwidth]{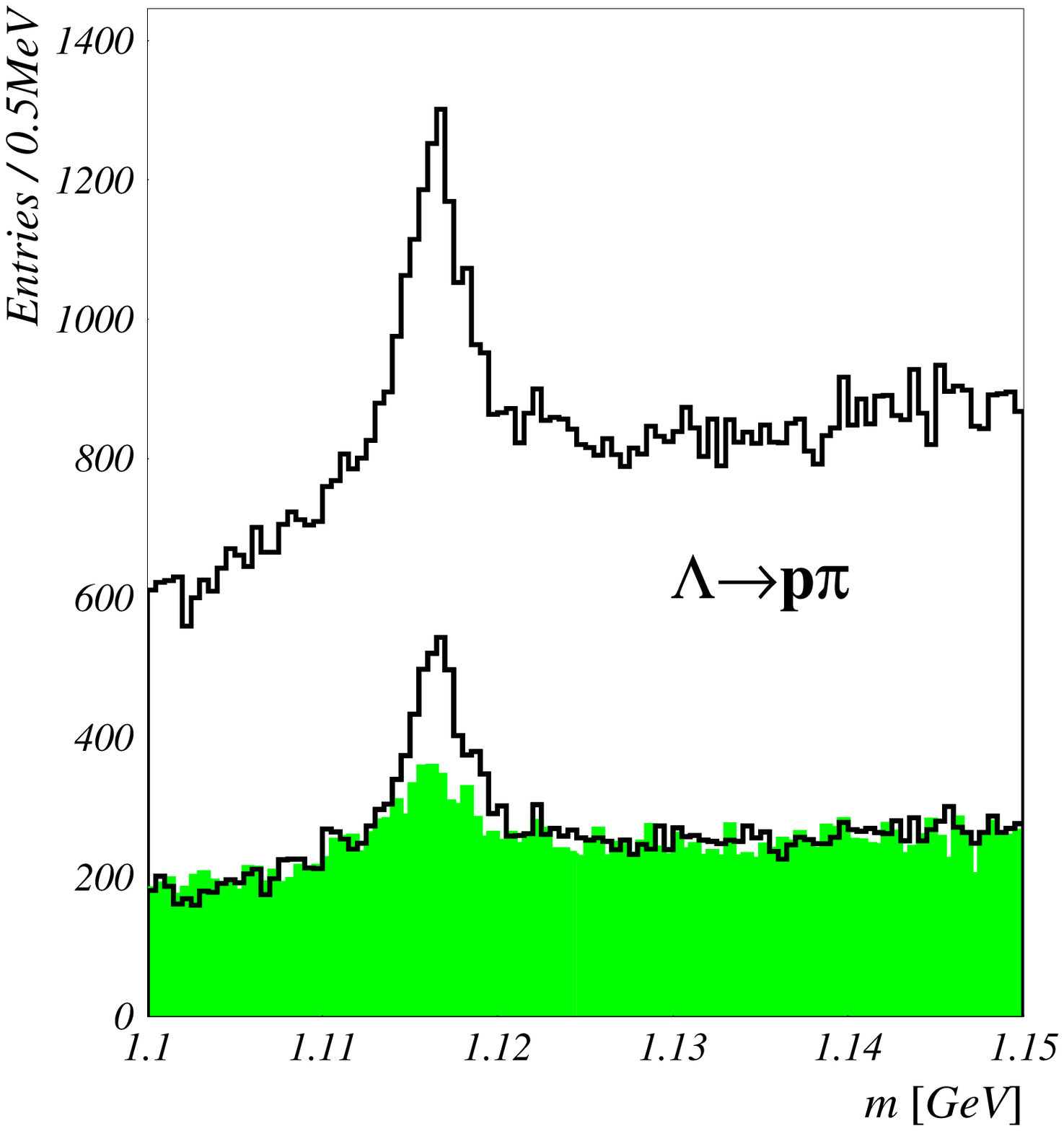}
\end{flushright}
\end{minipage}
  \caption[Comparison of reconstructed hadrons by MACRIB, RIBMEAN and HADSIGN.]
          {{\bf Left:} Comparison of reconstructed $\Phi$ mesons obtained by MACRIB (solid line)
            and the RIBMEAN routine (shaded area). The RIBMEAN result is compared to the signal given
            by MACRIB at the same background level (lower solid line) and at the same purity. At the
            same background level about 56\% more $\phi$ events are selected corresponding to a purity
            increase of 16\%. At the same purity the tagging efficiency is doubled.\\
            {\bf Right:} The comparison between MACRIB and HADSIGN
            in the $\Lambda\to p\pi$ decay. The solid line represents again the distribution
            by MACRIB and the shaded area comes from HADSIGN. For the distribution no cuts at all
            have been applied, in particular no decay length cut. At the same background level the
            efficiency to tag a $\Lambda$ is more than doubled. At the same purity the ratio is about
            4.}
  \label{fig:sample}
\end{figure}
The kaon network was tested on real data to observe the gain in efficiency in physically relevant
decays. The gain in efficiency is always dependent on the momentum region of the decay products,
as the networks are momentum dependent too, in their performance.

$\Phi$ mesons were reconstructed by identifying both kaons in the decay $\Phi \to K^{+} K^{-}$.
Using a cut of 0.8 in MACRIB's kaon network output for both tracks, about 56 \% more $\Phi$'s are
selected than with the combined RIBMEAN tight heavy tag at the same background level. The purity at
this cut is about 60\% for the signal obtained by the RIBMEAN tagging routine and approximately 70\%
in the distribution achieved by MACRIB (figure~\ref{fig:sample}~Left). The efficiency gain at the same
purity is thus much higher ($\approx 2$).

In the decay $\Lambda\to p\pi$ the output of HADSIGN has been compared to that of MACRIB.
At the same background level a large gain can be observed, in this case too
(figure~\ref{fig:sample}~Right). For the distribution
no cuts were applied at all, beside the proton tagging. This explains the relatively large background
level. At the same purity the tagging efficiency of MACRIB is four times as large as that of HADSIGN.

\begin{figure}[p]
  \centering
  \epsfig{file=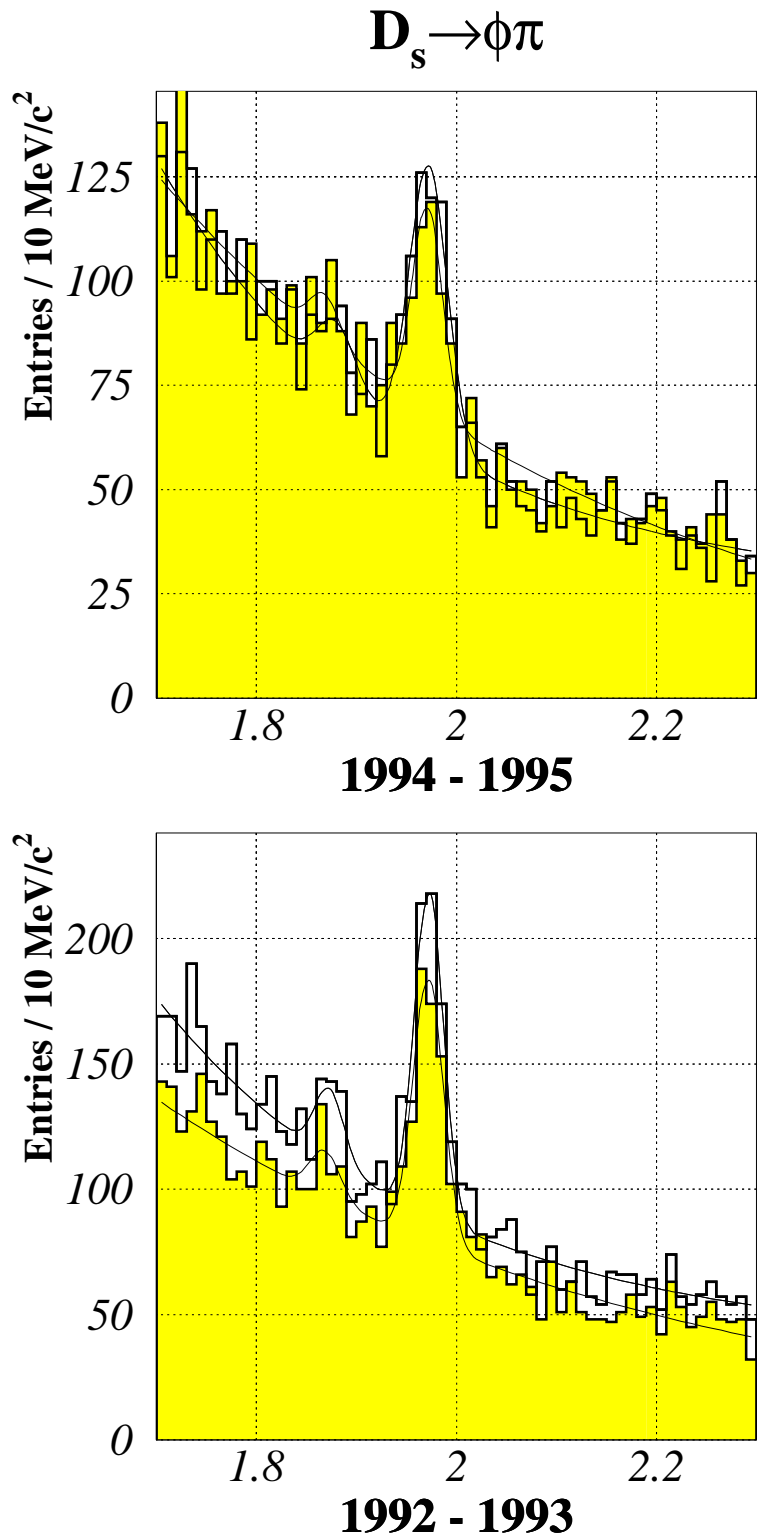,width=10cm
  ,clip=,bbllx=85,bblly=50 ,bburx=450,bbury=800}
  \caption{Comparison of reconstructed $D_s$ mesons decaying into $\phi\pi$ obtained by MACRIB (solid line)
    and the RIBMEAN routine (shaded area). The efficiency of tagging is in the case of MACRIB slightly
    better than in the case of RIBMEAN.}
  \label{fig:dsphipi}
\end{figure}

\begin{figure}[p]
  \centering
  \epsfig{file=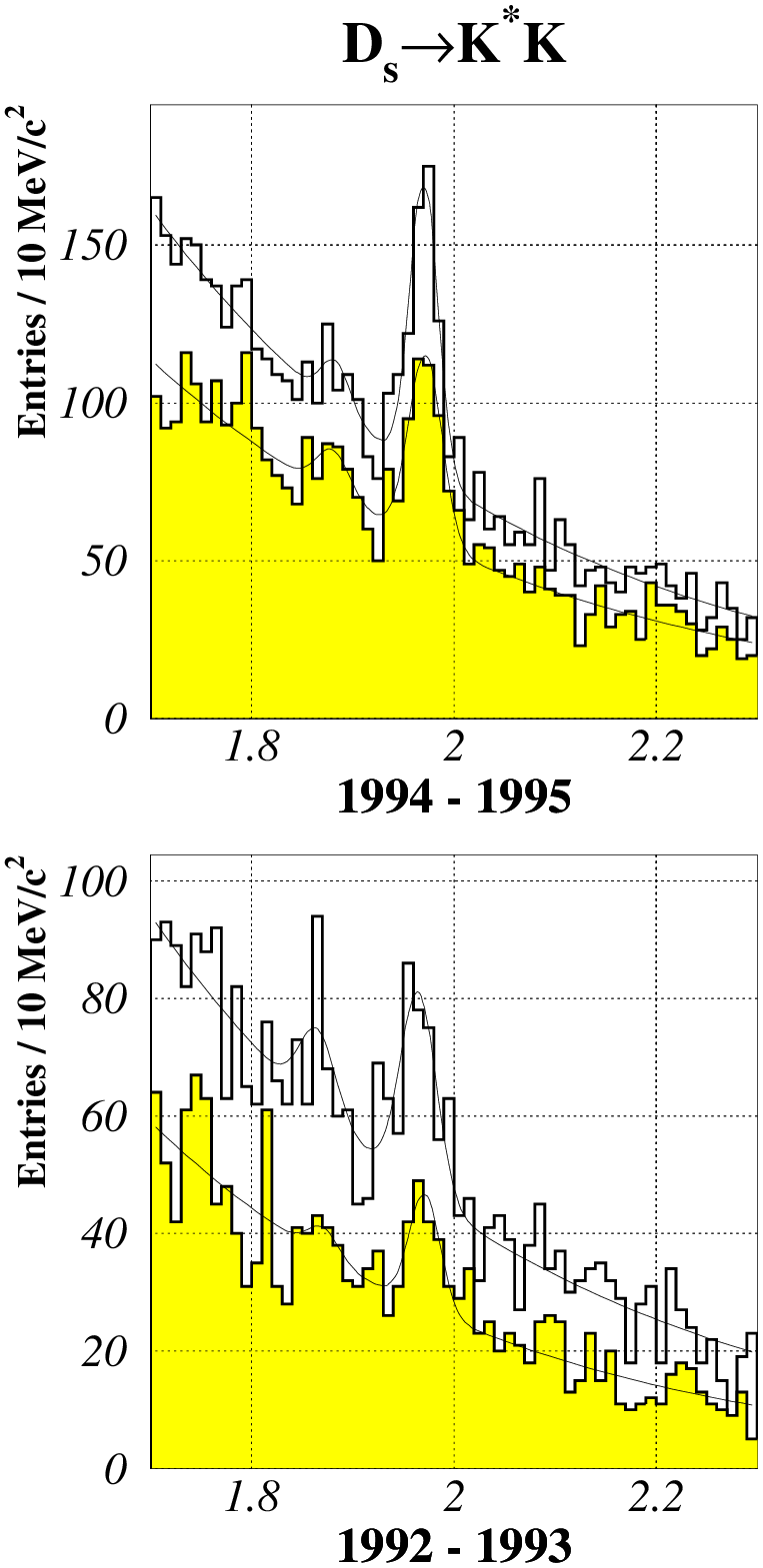,width=10cm,
    clip=,bbllx=85,bblly=50 ,bburx=450,bbury=800}
  \caption{Comparison of reconstructed $D_s$ mesons decaying into $K^*K$ obtained by MACRIB (solid line)
    and the RIBMEAN routine (shaded area). The plots show distributions at same purities. The efficiency gain
    for the data taken in the years 94-95 is about 33\% (above) and for the data taken in the years 92-93
    about 100\% (below).}
  \label{fig:dsksk}
\end{figure}

In the decay $D_s\to \phi\pi$ (figure \ref{fig:dsphipi}) RIBMEAN provides a comparable output
to that of MACRIB, since a very pure $\phi$ enrichment is possible without any kaon tagging. The decay
of the $D_s$ meson into $K^*K$ is shown in figure \ref{fig:dsksk}. It compares the output of MACRIB
and RIBMEAN at same purities for data taken in the years 1994-1995 and 1992-1993. The efficiency gain
is about 30\% and 100\% respectively.
\section{Acknowledgements}
Compared to the first version presented in the October 1998 DELPHI analysis
week, the net was still much improved, also due to implementation of feedback
obtained from RICH experts, especially Claire Bourdarios and Emile Schyns,
during that week. Many thanks to them.

We want to thank Jong Yi for providing the $\Phi$ and $D^*$ plots and 
Oleg Kuznets for providing the $D_s$ plots.
%

\end{document}